\documentclass{article}
\usepackage[T1]{fontenc}
\usepackage[utf8]{inputenc}

\usepackage{graphicx}
\usepackage{fullpage, amsmath, amssymb, amsthm}
\usepackage{caption}
\usepackage{subcaption}

\usepackage{physics,url}
\usepackage{graphicx}
\usepackage{subcaption}
\DeclareMathOperator*{\argmax}{arg\,max}
\newcommand{\fdm}{t_\textrm{mat}}

\title{Algorithms Clearly Beat Gamers at Quantum Moves\\A Verification}
\author{Allan Grønlund\\
Computer Science, Aarhus University, jallan@cs.au.dk}
\date{}
\begin{document}
\maketitle

\begin{abstract}
  The paper \cite{sorensen2016} considers how well human players compare to algorithms for solving the Quantum Moves game 'BringHomeWater' and design new algorithms based on the intuition extracted from the player solutions.  
  The claim by \cite{sorensen2016} is that humans outperform widely used algorithms, in particular an algorithm developed by \cite{sorensen2016}, based on the Krotov algorithm \cite{krotov} and that player intuition is crucial to develop new and better methods.
  However, as initially discussed by D. Sels \cite{Dries}, a standard Coordinate Ascent algorithm outperforms all player solutions 
from \cite{sorensen2016} by a very large margin. Albeit D. Sels \cite{Dries} only compare to player solutions, the simple algorithm also outperforms all algorithms based on player solutions and Krotov discussed by \cite{sorensen2016}, and it does so using much less time and iterations than the numbers presented in \cite{sorensen2016}.
In this paper we elaborate on the methods discussed by D. Sels \cite{Dries} and verify experimentally that the presented algorithm, Stochastic Ascent, solves the problem much better than any player and all the algorithms derived from player solutions in \cite{sorensen2016}. We also experimentally verify the theoretical analysis of the game presented by D. Sels, verifying that an approximate theoretically derived protocol solves the problem very well, much better than players for relevant durations,  and in fact behaves very similar to protocols found by algorithms. 
We add a comparison with gradient ascent, which in Quantum Control is known as the GRAPE algorithm \cite{GRAPE}. Starting from uniform random values this standard algorithm also outperforms all players by a large margin. GRAPE works at least as well as the methods developed in \cite{sorensen2016} that initialized with the best player solutions. A standard analysis of the solutions provided by GRAPE provides a starting point from GRAPE which  outperforms all the algorithms from \cite{sorensen2016} as well.  We also compare with our own basic Krotov algorithm, and show that the results of this algorithm is similar to the performance of GRAPE and also outperforms all players, and the Krotov based algorithm given in \cite{sorensen2016}. Repeating the analysis applied to solutions found by GRAPE, leads to an algorithm that easily outperforms all players and and algorithms developed in \cite{sorensen2016}.
%
These experiments verify and underline the results in \cite{Dries} that the conclusions made in \cite{sorensen2016} regarding algorithms and optimization are untenable. In fact the opposite conclusions are true. 
\end{abstract}

\section{Introduction}
The paper \cite{sorensen2016} presents and discusses results for a Quantum Moves 
game, {\it BringHomeWater}, where players have attempted to move a quantum state from one position to 
another in a simulated optical tweezers and atoms setup.
The paper compares the player solutions to a  specific numerical method the authors have tested. In particular, 
\cite{sorensen2016} show evidence that an algorithm, denoted KASS, based on the Krotov optimization 
method \cite{krotov} performs worse than solutions that the human players have 
come up with. Given the assumption that the Krotov method has been correctly 
applied, the evidence points to the fact that human players can outperform 
the Krotov method. This is the find and claim of \cite{sorensen2016} and it 
features prominent in the abstract of the paper. It leads the authors to 
conclude that ``players succeed where purely numerical optimization fails, 
and analysis of their solutions provide insights into the problem of optimization 
of a more profound and general nature.'' While it seems clear from the presented 
data that human players have indeed outperformed a particular implementation of the 
Krotov algorithm (and, according to 
\cite{sorensen2016}, as a consequence also algorithms like CRAB \cite{CRAB} and 
differential evolution approaches \cite{evolve}, which is stated to perform worse than Krotov), {\it this finding is of no significance.} 
In fact, as has been discussed first by D. Sels \cite{Dries}, a very simple approach using 
classical arguments can capture the {BringHomeWater} Quantum Moves game far better than the player approach. Furthermore, 
as also shown by \cite{Dries}, one of the simplest optimization algorithms available, 
Stochastic Ascent, can outperform all of the above. This clearly demonstrates that the 
problem solved in \cite{sorensen2016} is not hard at all and efficiently solvable 
using well-known numerical methods. The authors of \cite{sorensen2016} note
in the abstract that 
'Using player strategies, we have thus developed a few-parameter heuristic optimization 
method that efficiently outperforms the most prominent established numerical methods.'
This must be regarded as the major outcome of \cite{sorensen2016}. This is a 
completely misleading conclusion when a simple Stochastic Ascent algorithm 
can be used to solve the problem better than any of the methods discussed in \cite{sorensen2016}.  

%
In \cite{sorensen2016}, an implementation of Krotov named KASS is prominently featured and stated as the best algorithm found among several algorithm tested in \cite{sorensen2016}. It has a reported calculation time of 6 hours per run (with 2400 runs reported this is 600 days of calculation) so the first observation is that this is an extremely slow method. It is much heavier in computing time than any of the algorithms discussed in \cite{Dries, DriesarXiv}, and all the results we discuss. This leads us to conclude that the KASS method is seriously challenged by the problem at hand.
Surprisingly, there is no discussion in \cite{sorensen2016} of the standard gradient ascent approach, GRAPE \cite{GRAPE}. We show that GRAPE solves the Bring Home Water problem very efficiently, much better than players and KASS. Furthermore, we have implemented the Krotov algorithm  and find that it performs similarly to GRAPE. It is therefore very puzzling how the KASS algorithm \cite{sorensen2016} appears to be so poor.


\paragraph{BringHomeWater}
The BringHomeWater game is defined by a setup with two tweezers and an atom, and the job is to use the controllable tweezer to move the atom, initially caught by the other tweezer, to some fixed position.  The tweezer that has initially caught the atom cannot be controlled and does not change i.e. it does not move and is turned on at a fixed amplitude. The other tweezer is controllable by the user/algorithm, and may be move around and its amplitude is controllable as well. 
The job of a player or an algorithm is to use the controllable tweezer to move the atom from its initial position to a given target position, intuitively by moving the controllable tweezer towards the atom, and then transporting it back again to the target position, while ensuring that the atom does not escape the tweezer (spill over in BringHomeWater analogy) using as little time as possible. See \cite{bhw_game} for more details.

We note that, surprisingly, the numerical optimization problem that is solved in BringHomeWater  is not really defined in the paper \cite{sorensen2016}, neither in the main 
text, methods, or supplementary materials. In particular several important parameter values are not given. A description and the parameters can be found in \cite{Dries}, and we follow the description given there. Initially, the atom and the fixed tweezer is placed at position $x_\textrm{start} = 0.55$. 
The task is to move the atom to $x_\textrm{end} = -0.55$ using the controllable tweezer, that in the playable game is initially placed at 
$-0.55$ as well. 
The amplitude of the fixed tweezer is set to 130, and the controllable tweezer we allow to have an amplitude in the interval [0, 160]. 

The game is defined for any fixed  duration of time to move the atom. The larger the duration the easier the problem, and the goal is to solve the problem with the smallest possible duration.
Given a fixed time duration, a \emph{protocol} for the controllable tweezer is a list of positions and corresponding amplitudes of the controllable tweezer for a set of time-steps. The quality of a protocol is determined by the so called fidelity under the condition that we have zero velocity of the atom at the start and end of the protocol.
Denote the desired target state of the atom by $\bra \phi$ and the initial state as $\ket \psi$.
For a given protocol $P$, the state of the atom after the protocol is given by the time evolution operator $U_P$ applied to the initial state. The fidelity of the protocol $P$ is formally defined as
$$
F(P) =  |  \mel{\phi}{U_P}{\psi} | ^2
$$
which is what we want to maximize over choice of protocol $P$.
To be able to compute the fidelity we need to compute the initial and target state vectors and the required evolution operator $U_P$. This is specified in Appendix~\ref{appa}.

The fidelity  dictates whether the atom has  been correctly moved to the target state and the goal is to find solutions with a fidelity of at least 0.999 \cite{sorensen2016}. As in \cite{sorensen2016} we define the QSL of an algorithm as the smallest duration for which the algorithm has returned a protocol with fidelity 0.999. This gives one number to easily compare different methods for the problem.

\paragraph{Stochastic Ascent:}
To make the point clear that no elaborate algorithms or initialization is required for BringHomeWater, the Stochastic Ascent algorithm that solves BringHomeWater is given in Figure \ref{fig:cascent}. 
\begin{figure}[h]
  \begin{tabular}{l}
    \textbf{Stochastic Ascent Algorithm:}\\
    \hline 
    let $x_1,\dots, x_N$  be the tweezer position, amplitude pairs of the {\bf controlled} tweezer initialized randomly \\
    while not in local minima \\
    $\quad$ iterate over $x_i,$ for $i=1,\dots, N$ in uniform random order \\
    $\quad \quad \quad$ fix all parameters except $x_i$ and optimize over $x_i$ i.e. \\
    $\quad \quad \quad$ $x_i \leftarrow \argmax_{\hat{x_i}} \textrm{fidelity}(x_1,\dots,x_{i-1},\hat{x_i},x_{i+1},\dots,x_N) $  \\
    return $x_1,\dots, x_N$
\end{tabular}
  \caption{Optimization Algorithm for maximizing fidelity in BringHomeWater by \cite{Dries}.}
  \label{fig:cascent}
\end{figure}

This is a simple and straightforward Coordinate Ascent \cite{cord_asc} optimization algorithm from mathematical optimization.
The only thing that may be considered complicated, is the computation of the fidelity, but that has nothing do to with the algorithm itself.
In words, the algorithm is as follows. Let us say we are given 40 time steps to move the atom. Then we have 40 pairs of parameters, the position and amplitude, of the controllable tweezer at each of these steps. The task is now to find a set of parameters that makes the fidelity large.
First, initialize randomly the parameters. That will almost surely give a bad fidelity score. 
Now, repeatedly iterate over the parameters in random order. When the algorithm consider a parameter, the remaining 39 are fixed. The algorithm then picks the value, pair for the parameter considered, such that in combination with the fixed parameters, the fidelity is as large as possible.
Finally, these these random scans of the parameters are repeated until the fidelity stops improving.
You could visualize this as having 40 knobs you can turn up and down, and every time you turn one the score changes. The algorithm then simply randomly selects knobs to turn at random, and turns the knob chosen to the position maximizing the score with the remaining knobs fixed.
To solve the problem of computing the best value for a given parameter, D. Sels \cite{Dries} simply impose a uniformly spaced fixed grid of allowed values for the parameters (knobs)  and then simply tests all possibilities in the algorithm, instead of elaborately trying to find the exact best position, amplitude pairing.

\paragraph{Our Results:}
In this paper we analyze the results from \cite{sorensen2016} more thoroughly and verify the results presented by D. Sels \cite{Dries}, and then add comparison to basic  GRAPE and Krotov algorithms. We experimentally verify the results achieved with stochastic ascent, by making our own implementation of the algorithm and performing the same experiments as in \cite{Dries}. We also test a few variations of the algorithm. Our experiments match and verify the experiments given in \cite{Dries} that show that this simple algorithm easily outperform all player solutions and algorithms from \cite{sorensen2016}, and that it is indeed very fast, much faster than the algorithms presented in \cite{sorensen2016}.
We also experimentally test the theoretical work performed by D. Sels \cite{Dries} and verify that the fidelity reported for the theoretically derived protocol are correct.
We then compare the excitation of the atom to be transported in the theoretically derived protocol to the excitation of the atom caused by the  protocol found by stochastic ascent and  show that they actually look quite similar at the low energy levels.  Finally, we use the theoretically derived protocol as a starting point for the GRAPE algorithm and this combination easily outperform all players and all algorithms developed in \cite{sorensen2016} in one try. This implies that the theoretically derived counterdiabatic approximate protocol is in fact quite good.

Now Stochastic Ascent is a standard Coordinate Ascent algorithm algorithm. It is, however, not as famous as as the gradient ascent algorithm. To add perspective to the algorithmic part of the problem, we test the more famous gradient ascent algorithm which in Quantum Control is know as the GRAPE algorithm described in 2005 \cite{GRAPE}.
First, we initialize GRAPE with completely uniform values for both positions and amplitudes. With this basic setup GRAPE clearly outperforms all players and the KASS algorithm. This is highly surprising since gradient ascent is perhaps the most famous algorithm in mathematical optimization and has been considered in Quantum Control as the GRAPE algorithm for more than a decade. Furthermore, we started the algorithm from the most naive starting positions and amplitudes imaginable. By visualizing the positions and amplitudes output by the GRAPE algorithm with heat maps, we extract new starting positions and amplitudes which when used with the GRAPE algorithm, clearly outperforms all players and all algorithms developed in \cite{sorensen2016}.
%
We continue to investigate the random starting positions and amplitudes used in the KASS algorithm. Since the KASS algorithm is not fully specified in \cite{sorensen2016} we do a best approximation. We run the GRAPE algorithm from these random starting positions and amplitudes  and show that the GRAPE algorithm with these initial values also outperforms all human players and the actual KASS algorithm by a wide margin.

While the KASS algorithm is a special version of the Krotov algorithm with specific seeds and sweeps, it is surprising that the Krotov algorithm, a standard algorithm in Quantum Control, performs so much worse than GRAPE and Stochastic Ascent. We implement a standard Krotov method and apply it in the same way as GRAPE with completely random starting positions and amplitudes.
The results for our  Krotov implementation resemble those achieved with GRAPE, and easily outperform all players and the KASS algorithm. Comparing to \cite{sorensen2016}, both our GRAPE and Krotov implementation  take only a thousand iterations over the parameters to converge, nothing like the more than 100 million  trials as stated in \cite{sorensen2016}.

\begin{figure}[ht]
  \centering
  \includegraphics[width = 0.8\columnwidth]{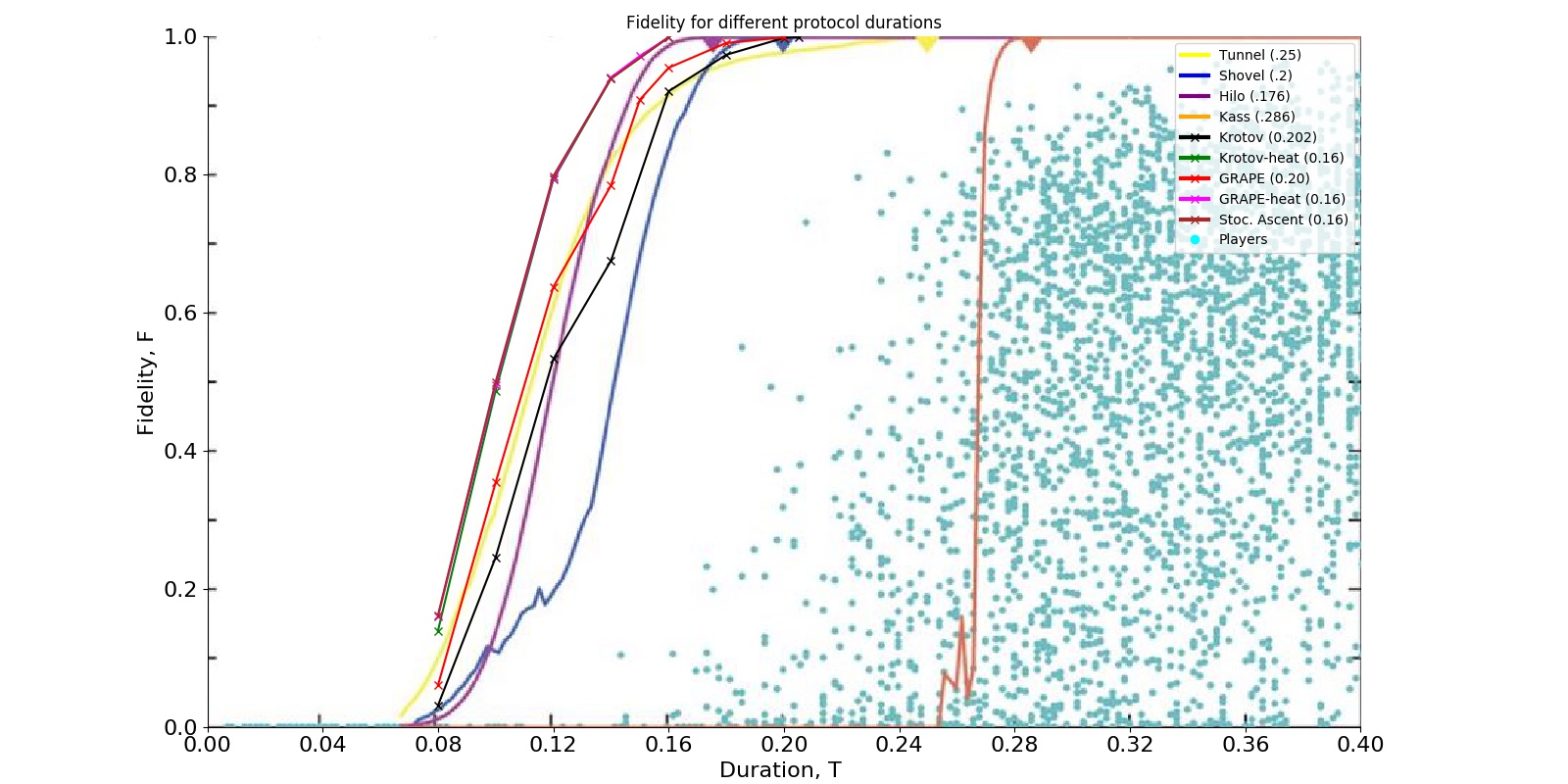}  
  \caption{Fidelity for different BringHomeWater solutions as a function of duration $T$.
    Each algorithm Quantum Speed Limit (min duration $T$ with fidelity above 0.999) is shown in parentheses.
    The -heat results are the same algorithm initialized with the protocol and amplitude found by extracting from a heat map of positions and amplitudes of several runs. We emphasize, as it may be hard to see, that the results for Stochastic Ascent and the heat solutions are nearly identical.
  }
  \label{fig:all_comparison}
\end{figure}

We have shown the results achieved for different durations by our standard implementation of standard algorithms in Figure \ref{fig:all_comparison}.
For each duration the goal is to have as high as fidelity as possible and the longer the duration the easier the problem. Thus the goal is to be close to the left top corner. As we can see from the figure all the algorithms behave fairly similar except for the KASS algorithm. It is extremely poor when compared with the other basic algorithms, including our implementation of the Krotov algorithm.  This proves that the players do not succeed where algorithms fail nor do player inspired solution outperform even basic algorithms in any way, which was the main conclusions from \cite{sorensen2016}. In fact the figure show the opposite to be true.

\paragraph{Outline}
In Section \ref{sec:stoc_ascent} we present all our experiments with stochastic ascent. In Section \ref{sec:grape} we present our experiments and results for the GRAPE algorithm, and in Section \ref{sec:krotov} we show our experimental results for the Krotov algorithm. In Section \ref{sec:cd_theory} we provide our experiments with the theoretical derived protocol from \cite{Dries}.
Finally, we summarize in Section \ref{sec:conclusion}. The appendix contains all the needed derivations for our algorithms and code for Stochastic Ascent, GRAPE and Krotov is available at GitHub.

\section{Stochastic Ascent Experiments}
\label{sec:stoc_ascent}
In this section we experimentally show that the Stochastic Ascent algorithm as presented in \cite{Dries} is far superior to human players and the algorithms presented in \cite{sorensen2016} that use player solutions as initial seeds for other algorithms. 

In all experiments, the amplitude of the controllable tweezer is set to 160 if nothing else is stated following \cite{Dries}. 
Extended  Data Figure 3 of \cite{sorensen2016} indicates that an amplitude of $A\sim 150$ may be the best of highest value  for the amplitude of the controllable tweezer since this is the maximal value used by the algorithms in  \cite{sorensen2016}, that states in regards to KASS that `Optimizations showed that
$A=-150$ was the optimal value'.
This seems to mean that the 150 is the best value to use for the algorithms used in \cite{sorensen2016}. Here we use the values provided by \cite{Dries} taken from the BringHomeWater game itself, i.e. we use the maximal allowable amplitude as provided to the gamers.

The reason for fixing the amplitude at the maximal value is based on two observations. First, the analytic results in \cite{Dries}  shows  consistently better results when the amplitude is larger.   Second, since it is a linear control field in the hamiltonian, Pontryagin theorem tells us that the solution is (almost) always on the boundary. This means that $A=160$ or $A=0$ at any time and you can switch between these two. But since $A=0$ means that the atom is not being confined the most likely optimal solution is that $A=160$ all the time (besides maybe for the first step).
For completeness and generality,  we also tested adding more allowable amplitudes, which is simple extension of the algorithm, but that only made the algorithm slower without changing the results, supporting the theory and the choice in \cite{Dries}.

To cover the different possibilities we consider different versions of the Stochastic Ascent algorithm from \cite{Dries}. We add a version where the first position of the controllable tweezer is always $-0.55$ as in the game, and a version where additionally the amplitude of the controllable tweezer is set to 150.

\subsection{Full Comparison}
\label{sec:full_cmp}

\begin{figure}[ht]
  \centering
  \includegraphics[width = 0.8\columnwidth]{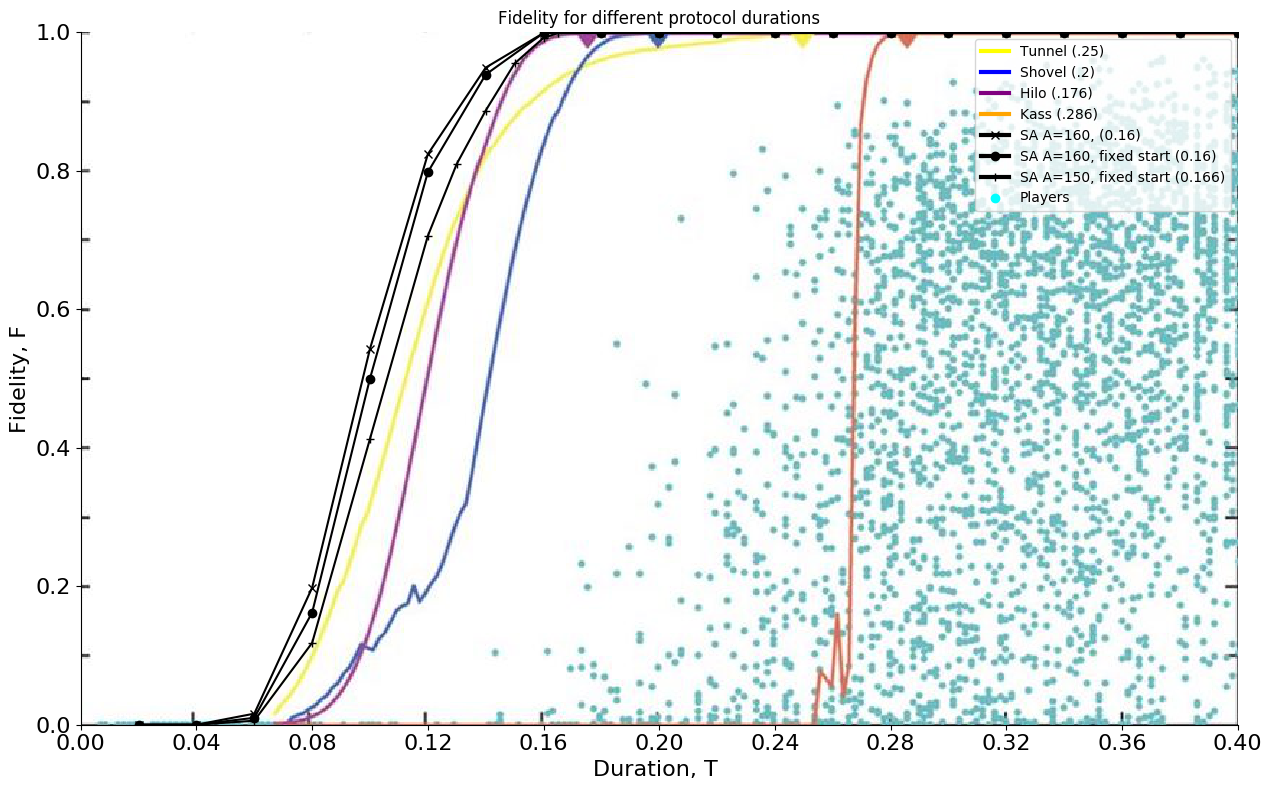}  
  \caption{Fidelity for different BringHomeWater solutions as a function of duration $T$.
    Each algorithm Quantum Speed Limit (min duration $T$ with fidelity above 0.999) is shown in parentheses.
    Fixed start means that the controllable tweezers first position is fixed to $-0.55$ for the Stochastic Ascent.
    For all Duration $T$ tested, the number of steps $N$ is defined such that $T/N=0.0025$ as in Figure 6 in \cite{Dries}.
    We use 201 possible position for the controllable tweezer which allow putting the controllable tweezer at $x_\textrm{start}=-0.55$ as required.
  }
  \label{fig:comparison}
\end{figure}


In Figure \ref{fig:comparison} we have shown a comparison of the Stochastic Ascent algorithms  with the players and the derived algorithm from \cite{sorensen2016}. For each duration $T$ we can compare the score of the different approaches, higher being better. This plot clearly shows the superiority of the simple Stochastic Ascent algorithm over all players by a massive margin. The gamers are not even close to the performance of the simple algorithm. 
Comparing with the algorithms derived by using player solutions and the KASS algorithm, the plot shows that these are also clearly beaten by the simple algorithm.

The difference between fixing the first position of the controllable tweezer at $-0.55$ and allowing the algorithm to optimize this step is very small and makes essentially no difference for the result.
If we reduce the amplitude of the controllable tweezer to 150 the results are also slightly worse. This reduction in QSL is predicted by \cite{Dries} that says QSL should follow $1/\sqrt{A}$ and the reduction we see in the experiments fits well with that. If we allow a higher $T/N$ ratio and and tune other parameters like the number of possible tweezer positions the QSL may possibly be reduced for all versions of the Stochastic Ascent algorithm. Since the results of the Stochastic Ascent are already clearly superior, and thus verifying \cite{Dries}, we do not follow this direction further in this note.

To experimentally verify that it is a good strategy to start by a fast move from the starting position at $-0.55$ to where the atom is caught by the fixed tweezer we have visualized a protocol for $T=0.16$ that achieves a fidelity above $0.999$.
\begin{figure}[ht]
  \centering
\includegraphics[width = 0.6\columnwidth]{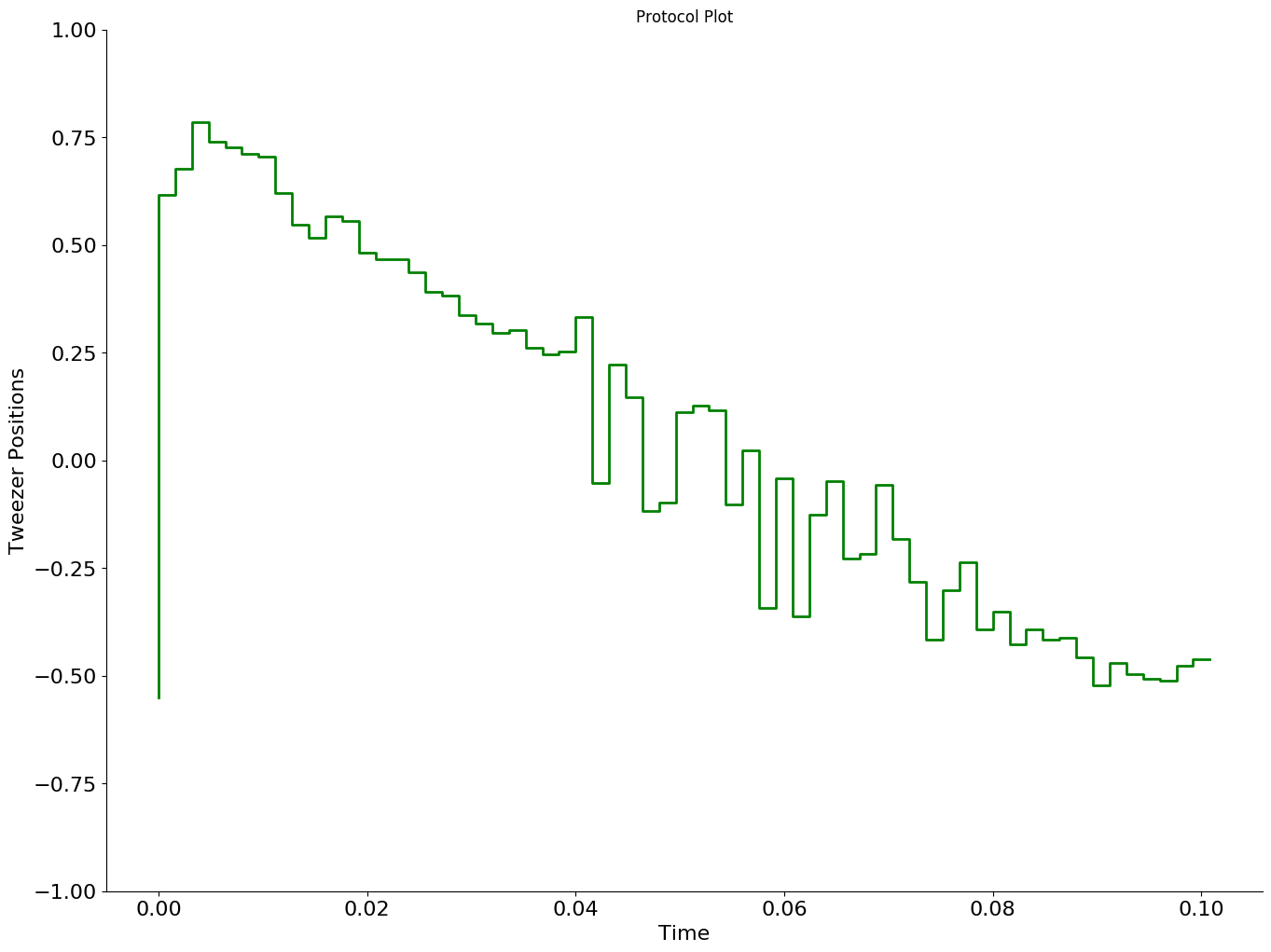}
\caption{Tweezer positions for a protocol with fidelity $> 0.999$ at duration $0.16$, steps 64, with fixed starting position at $-0.55$. }
  \label{fig:SDposition}
\end{figure}

\subsection{Fidelity Traces}
\begin{table}[ht]
  \centering
\begin{tabular}{cc}
  \begin{subfigure}[t]{0.49\textwidth}\centering\includegraphics[width=1\columnwidth]{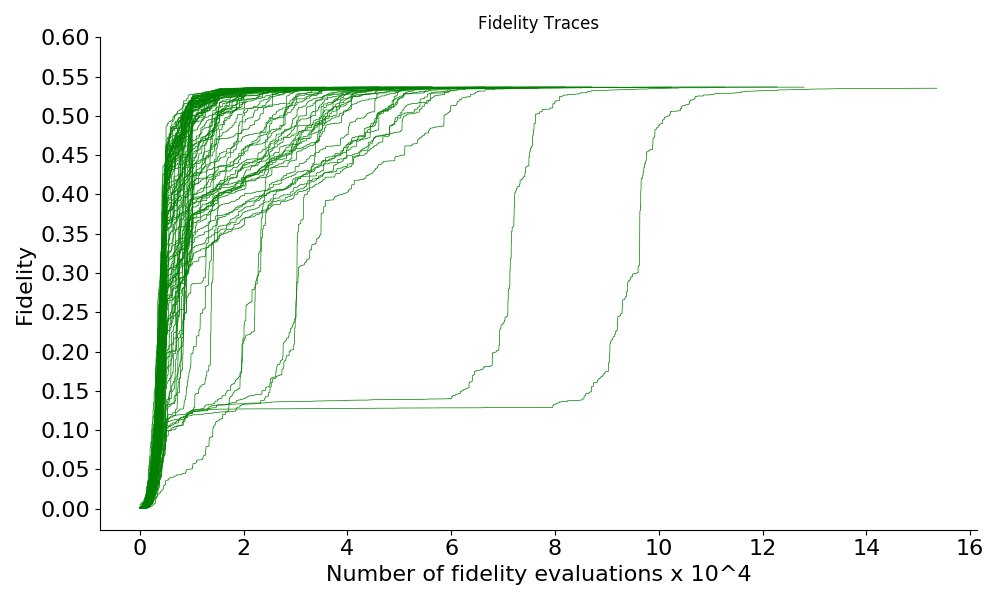}
    \caption{Fidelity Trace, duration 0.1, 40 steps, 128 allowed positions for the controllable tweezer, 100 traces as in \cite{Dries}.
      Fidelity range of experiment, 0.533 -  0.537
    }
    \label{fig:taba}\end{subfigure}
  &
  \begin{subfigure}[t]{0.49\textwidth}\centering\includegraphics[width=1\columnwidth]{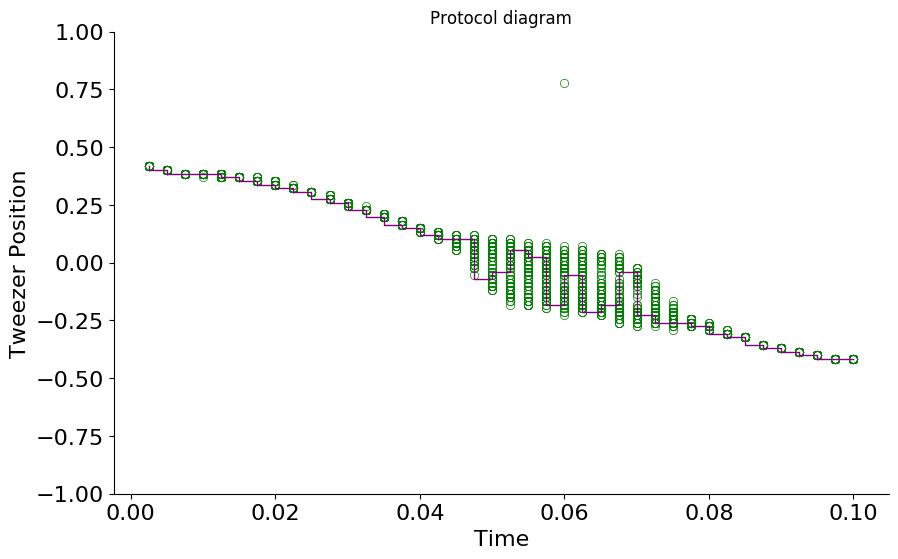}
    \caption{Final Protocol Scatter Plot of the traces from \ref{fig:taba}. A single protocol has been outlined in purple.}
    \label{fig:tabb}\end{subfigure}\\
\newline
\begin{subfigure}[t]{0.49\textwidth}\centering\includegraphics[width=1\columnwidth]{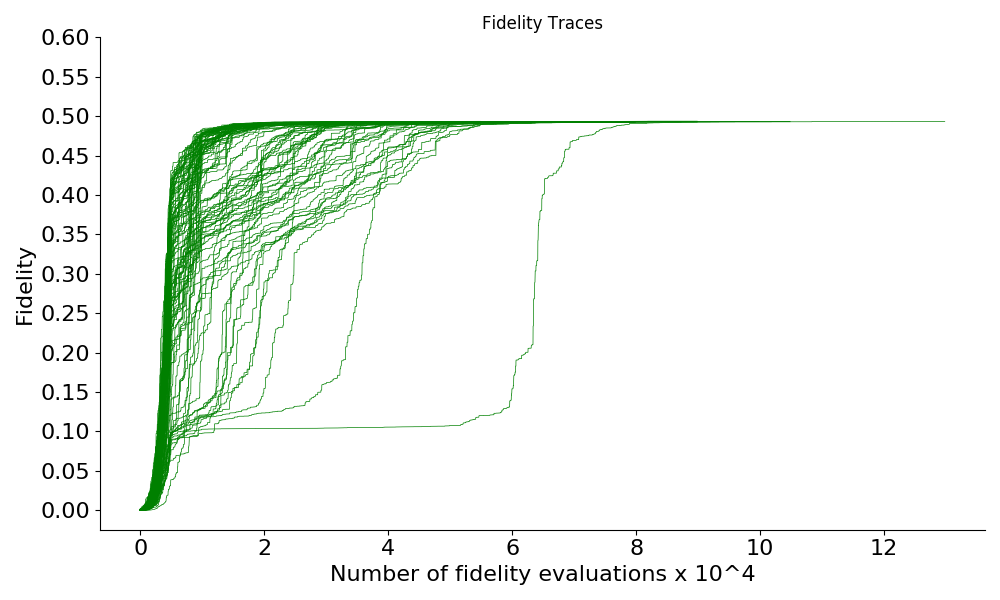}
  \caption{Fidelity Trace, duration 0.1, 40 steps, 128 allowed positions for the controllable tweezer, 100 traces as in \cite{Dries} but with first position of controllable tweezer fixed to (almost) $-0.55$.
    Fidelity range of experiment 0.489 - 0.493.
  }
  \label{fig:tabc}\end{subfigure}
&
\begin{subfigure}[t]{0.49\textwidth}\centering\includegraphics[width=1\columnwidth]{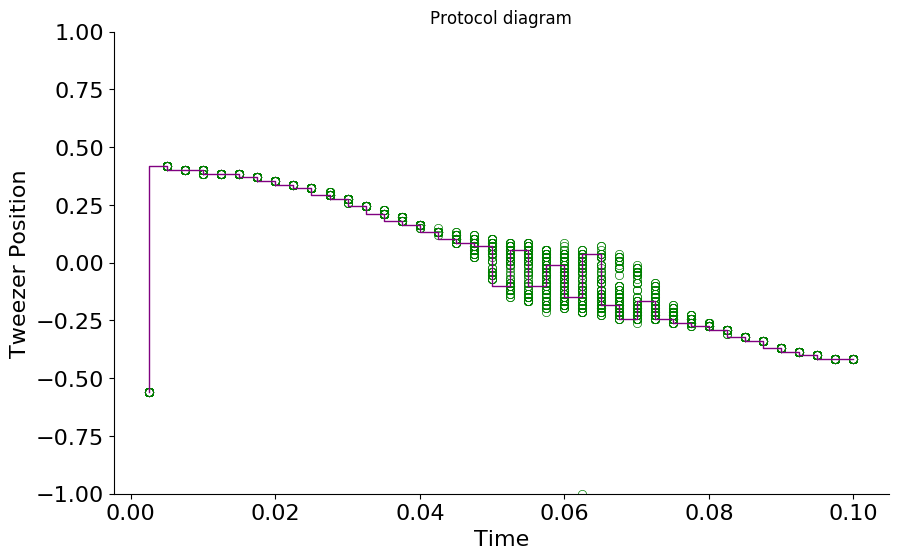}
  \caption{Final Protocol Scatter Plot of the traces from \ref{fig:tabc}. A single protocol has been outlined in purple.}
  \label{fig:tabd}\end{subfigure}\\
\end{tabular}
\caption{The fidelity trace and protocol plot as shown \cite{Dries}
(note that the protocol plot is in the arXiv version \cite{DriesarXiv} and that the amplitude of the controllable tweezer is 160)}
\label{tab:fid_traces}
\end{table}

We have also created the fidelity trace plot from (\cite{DriesarXiv} Figure 6), that zooms in on one the most volatile durations at $T=0.1$. 
We have created the plot for the two invocations of the algorithm as above, one invocation where the first position is optimized in the algorithm and one invocation where the first position is fixed at $-0.55$,  see the figures in Table \ref{tab:fid_traces}. A simple visual comparison shows that our implementation gets results that are essentially identical to the results reported in \cite{DriesarXiv}.

The simple conclusion is that the algorithm essentially gets the same fidelity all the time with different but very similar protocols having found local minima (in the sense that one parameter change cannot increase the score) with almost the same fidelity. As noted in \cite{Dries} the algorithm converges after very few iterations over the parameters, each round taking  only a few seconds on a laptop. 
Fixing the first position of the tweezer does not change this, only reduces the fidelity achieved slightly for this duration as was also apparent from Figure \ref{fig:comparison}. 
It is not surprising that fixing the first position of the controllable tweezer does not change much, since fixing the first position essentially just corresponds to a slight perturbation of the starting state from $\ket \phi$ to  $U_{-0.55} \ket \psi$, where $U_{-0.55}$ is the evolution operator the first step with the controllable tweezer at $-0.55$, and then running the algorithm from this configuration instead. 
The simple algorithm does not care about the starting and ending state, and if we add the fact that we could have used the smallest possible amplitude allowed for the controllable tweezer for that position of the protocol the change in the starting state would be even smaller. In the extreme case of the amplitude of the controllable tweezer being zero at the first step the only force acting on the atom is the fixed tweezer that already has the atom caught. 
The difference in fidelity between the two algorithms in Table \ref{tab:fid_traces} is around 0.04. 
Running Stochastic Ascent while not fixing the first position for a duration $T=0.0975$ and $N=39$ (maintain the same $T/N$ trade off) the fidelity achieved hovers closely around 0.492, essentially the same as for Stochastic Ascent with fixed first position with one additional step and time unit.
As the fidelity achievable varies greatly around $T=0.1$ losing a step means losing $T/N$ time which matches the difference between the two versions explaining the discrepancy. 
Since we have explained the difference we see no reason to include an experiment with a fixed starting point for the controllable tweezer at $-0.55$ while allowing amplitude of zero.
We conclude that our experiments are in full agreement with the results reported by D. Sels \cite{Dries} verifying his results.
Note that the protocols that are found by both version of the algorithms are the simple intuitive ones as above: Move the tweezer to the atom, move the tweezer back slowly, and shake it on the way to avoid spilling.

\subsection{Superposition states}
As discussed in \cite{Dries}, the BringHomeWater (Quantum Moves) may not be the best representation of quantum problems.
To add as much quantum complexity as possible to the BringHomeWater challenge, \cite{Dries} suggests replacing the target state $\bra \phi$ with a superposition of the target state and end the initial state, i.e.
$\bra \phi_{q_1,q_2} = q_1 \bra \phi + q_2 \bra \psi$. In Table \ref{tab:super_protocols} we have the fidelity achieved for different durations and a representative high fidelity protocol for a specific superposition. Now this protocol, looks much more complex than in the ``vanilla'' BringHomeWater and it would be fun to see if human players can get high fidelity here (as the simple stochastic ascent algorithm can).
\begin{table}[ht]  
  \label{fig:super_prototols}
  \centering
    \begin{subfigure}{0.4\textwidth}\centering
      \centering \includegraphics[width=1\columnwidth]{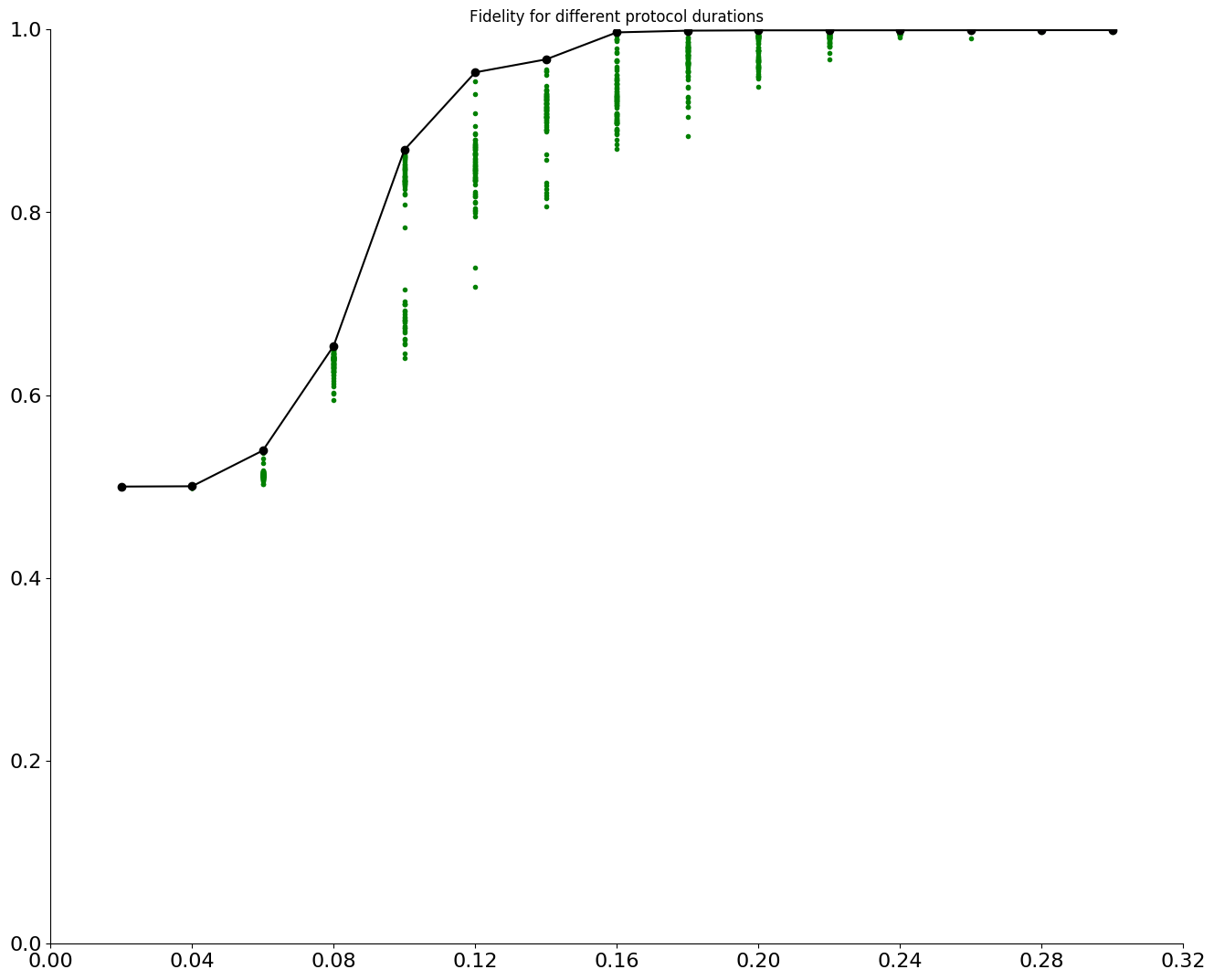}
      \caption{Fidelity of superposition $\sqrt{2}/2 \bra \phi + i \sqrt{2}/2 \ket \psi$ at different durations (50 repeats per duration). Green dots are different runs,  black circles show the max fidelity achieved.}      
    \label{fig:no_phase}
    \end{subfigure}\quad    
    \begin{subfigure}{0.4\textwidth}\centering
      \includegraphics[width=1\columnwidth]{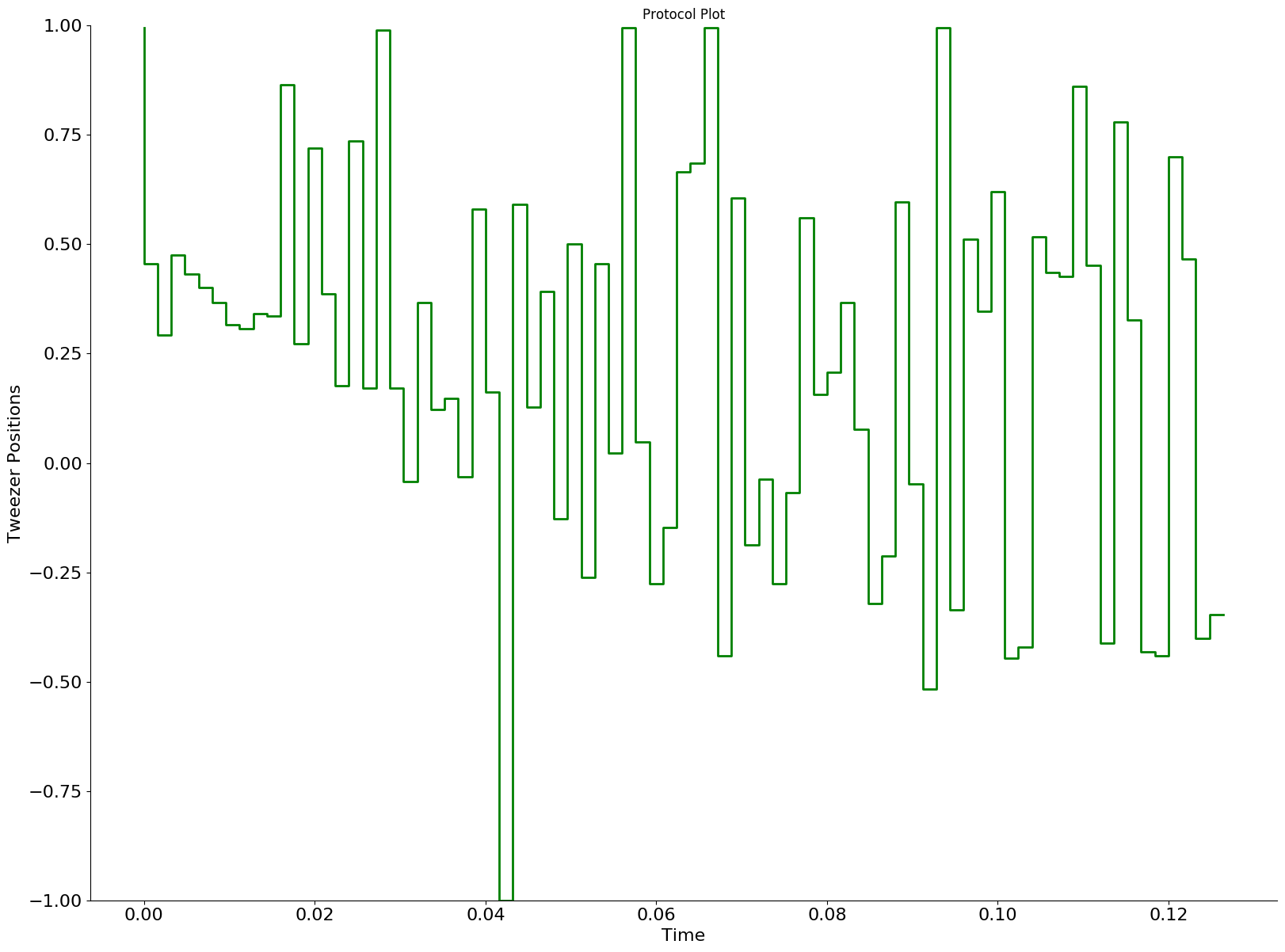}
      \caption{Protocol for target state: $\sqrt{2}/2 \bra \phi + i \sqrt{2}/2 \ket \psi$, duration 0.2, 80 steps with high fidelity ($\geq 0.999$)}
      \label{fig:phase}
    \end{subfigure}
    \caption{Results for Superposition Target State}
    \label{tab:super_protocols}
\end{table}
The Stochastic Ascent algorithm is able to solve the superposition version of the problem as well, albeit as noted in \cite{Dries} the variance of the result of a run is higher for this problem than it was for the vanilla BringHomeWater, and more runs of the algorithm may be required to achieve high enough fidelity. This indicates that this is more complex optimization problem. 

Together with the previous experiments this shows that the simple algorithm efficiently solves the BringHomeWater game in any configuration without any prior information with better performance than all results provided in \cite{sorensen2016}.

\section{Gradient Ascent Algorithms}
\label{sec:grape}
Seeing that a standard algorithm from mathematical optimization completely outperforms all players and the best results in \cite{sorensen2016} that has used more than ten thousand computer hours raises many questions. For instance, how poor is this KASS algorithm which is stated to be the best algorithm the authors found for BringHomeWater in \cite{sorensen2016}. Since we already know how to compute the fidelity of a given protocol  we can straightforwardly implement a gradient ascent algorithm, which in quantum control is known as the GRAPE algorithm.  For this case, GRAPE is no more than gradient ascent on the tweezer positions and amplitudes. So instead of fixing a grid of allowed positions and amplitudes and iteratively improving the cost by picking one variable pair to optimize, GRAPE instead computes the gradient of the fidelity of all positions and amplitudes and takes a step along that direction. As mentioned earlier, this algorithm is surprisingly not mentioned  \cite{sorensen2016} despite that it appears that it is a standard algorithm in quantum control \cite{GRAPE}.  The implementation details can be found in Appendix \ref{appa}, in short we implement the fidelity computation in  TensorFlow and use automatic differentiation.

We have not elaborately tested gradient ascent, since there are so many variants and tools to improve on the quality and convergence of the methods. We just want to check how a normal implementation compare with the other solutions. The main question we need to consider is how to initialize the protocol before starting the iterative algorithm.
Gradient ascent requires that we give the algorithm a reasonable starting point. Finding a bad one is not hard, for instance fixing all the initial tweezer positions at the left end or something similar as a starting position gives essentially zero gradient to work with and  the gradient ascent algorithm fails to do anything.  Gradient ascent does not search the space of solutions in some clever way, the algorithms moves uphill until it dies down in a flat area, so if we start it at a poor place the results will be poor. No surprise there.  
To hinder the algorithm as much as possible, without killing it, we start by testing GRAPE using the classic and very poor strategy of assuming that we do not know anything about the problem and hence do uniform sampling of parameters in their range. This means that the positions are sampled uniformly at random from $[-1, 1]$ and the amplitudes are chosen uniformly at random in $[0, 160]$.

\begin{figure}[ht]
\centering \includegraphics[width = 0.7\columnwidth]{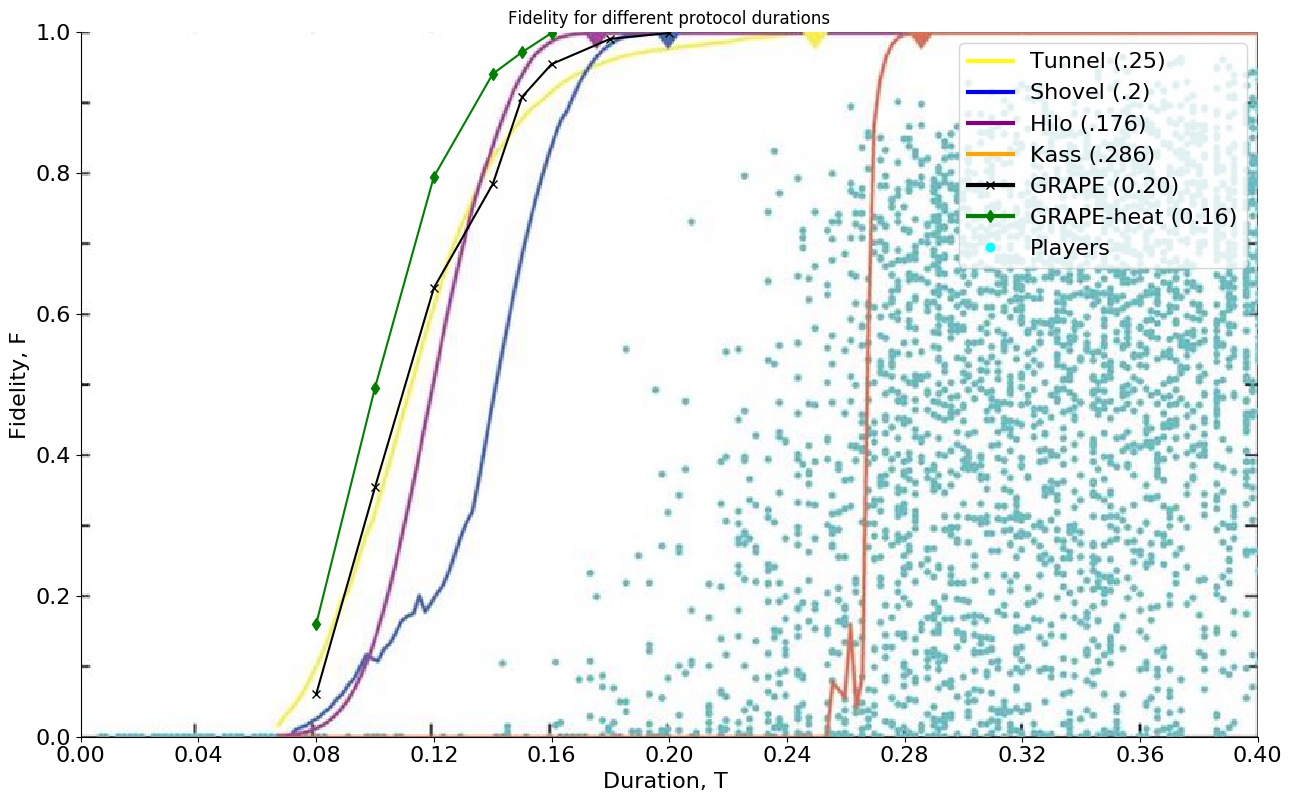}
\caption{Comparing GRAPE with results from \cite{sorensen2016}.
  GRAPE is gradient ascent with uniform initialization of tweezer positions and amplitudes.
  GRAPE-heat is gradient ascent initialized with the heat-map solution of 2000 GRAPE runs with uniform initialization.}
  \label{fig:grape}
\end{figure}
The results for the uniform initialization out-of-the-box GRAPE implementation  is shown in Figure \ref{fig:grape}.
From this figure it is obvious that GRAPE also drastically outperforms all players and the KASS algorithm by a wide margin.
In fact, GRAPE with completely uniform initialization is as good or better than running the algorithm from  \cite{sorensen2016} on the players seeds.
An if we use the protocol found for 0.18 that gets a fidelity of 0.99, just shy of the mark, as input for duration 0.19, we get a fidelity of 0.999, making it clearly better than starting from player seeds with the algorithm in \cite{sorensen2016}. Still, this GRAPE with completely uniform initialization is slightly worse than the HILO algorithm derived from analyzing player solutions in \cite{sorensen2016}.


\subsection{Intrinsic Intuition of Gradient Ascent}
Now if we do not wish to rely on any theoretical or intuitive understanding of the how to move the tweezer to get a better starting point than uniformly at random, we can instead analyze the protocols found by GRAPE. This is what the authors in \cite{sorensen2016} do with the player solutions to obtain their best algorithm.
However, instead of trying an elaborate analysis of the protocols, applying different clustering and visualization techniques, we simple make heat map plots of the protocols,  ignoring any dependencies between steps. The heat maps are constructed by splitting the position domain, $[-1, 1]$, into 40 equal sized intervals,  and for each step counting how many times a protocol has a position value in the each of the 40 intervals. We do the same for the amplitudes, using 32 buckets. We have shown the resulting heat map for duration 0.12 and 0.18 in Figure \ref{fig:heat_map}. In the Appendix we include the heat maps for the durations (0.1, 0.14, 0.16, 0.20) in Table \ref{tab:heat_maps}.  Inspecting the heat maps, the first thing that comes to mind is that the amplitude is almost always the maximal value (the top white bar), and in the few cases the amplitude does not converge to max, it is zero. Just like the theory suggested!
The only difference is step one, where the controllable tweezer is far from the atom, here it seems that there is no preference and it has essentially no impact on the result.
For the position of the controllable tweezer, the plots reveal the natural approach of moving the tweezer to the atom fast (in one step) and then slowly back again, showing that gradient ascent find this solution without knowing the game design.

It also clear from the heat-maps that from small durations (0.1-0.14) good GRAPE solutions all look almost exactly the same while for larger duration (0.16-0.18) it seems several solutions path exist among the best solutions found by GRAPE and as we increase the duration (0.20)  most of the good solutions again look similar but different from the small duration protocols.
Using this we create a new GRAPE experiment where we trace out the solution implied by the heat map for both amplitude and position. 
For the heat map solution for duration 0.16 we traced out the two most prominent paths and used as starting positions.
The results is also shown in Table \ref{fig:grape}, where we can clearly see that with this initialization easily extracted from the heat map, outperforms all algorithms from \cite{sorensen2016}.
In fact we get results just as good as we did for Stochastic Ascent albeit it took a little longer time to get there.

In conclusion, GRAPE with no information explores a much larger portion of the space because of the uniform initialization and still learns the intuitive protocols without access to first learning from simpler versions of the game nor the description or the graphical visualization of the task and gives better results than players and all algorithms from \cite{sorensen2016}.

\begin{figure}
  \begin{subfigure}[t]{0.5\textwidth}
    \centering
    \includegraphics[width=.4\linewidth]{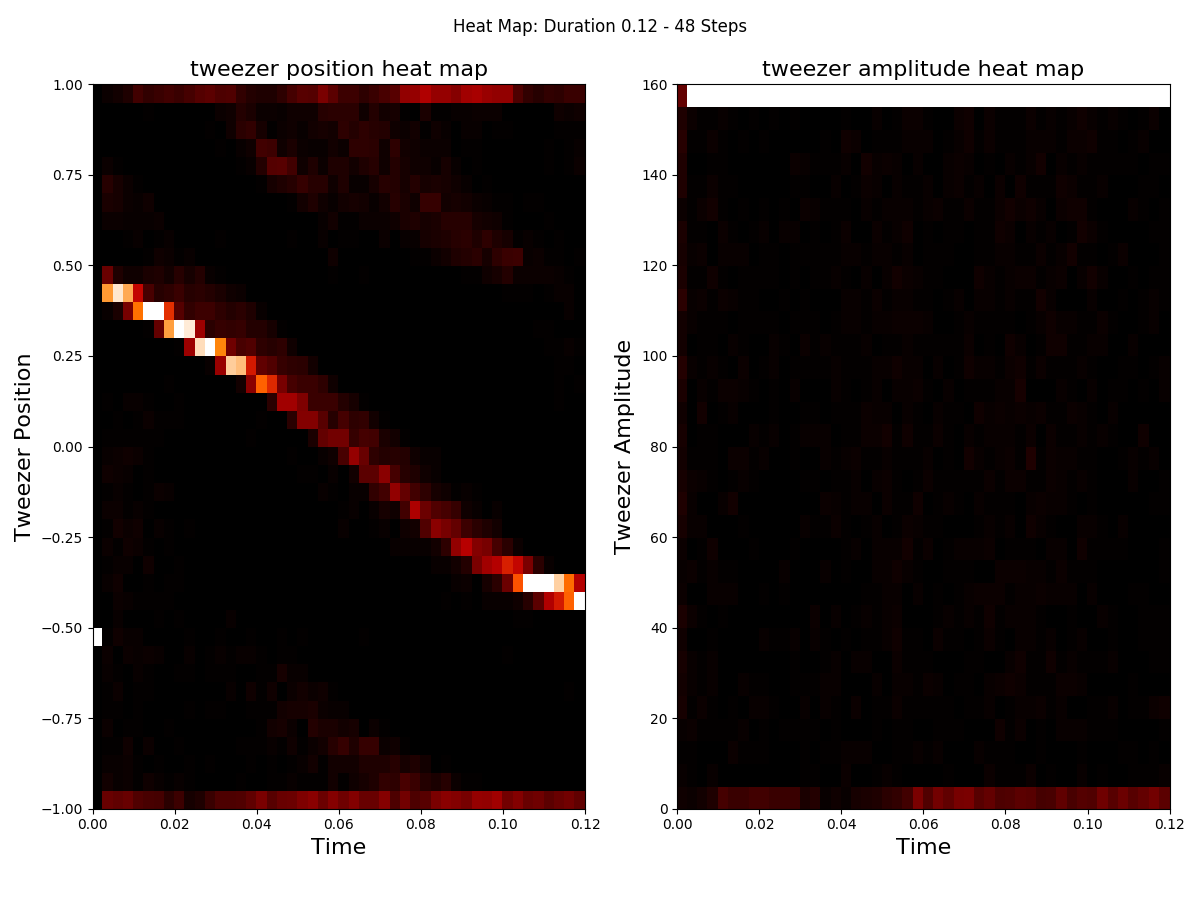}
    \caption{Grape Position and Amplitude Heat Map for duration 0.12, using 48 steps}
  \end{subfigure}%
  \begin{subfigure}[t]{0.5\textwidth}
    \centering
    \includegraphics[width=.4\linewidth]{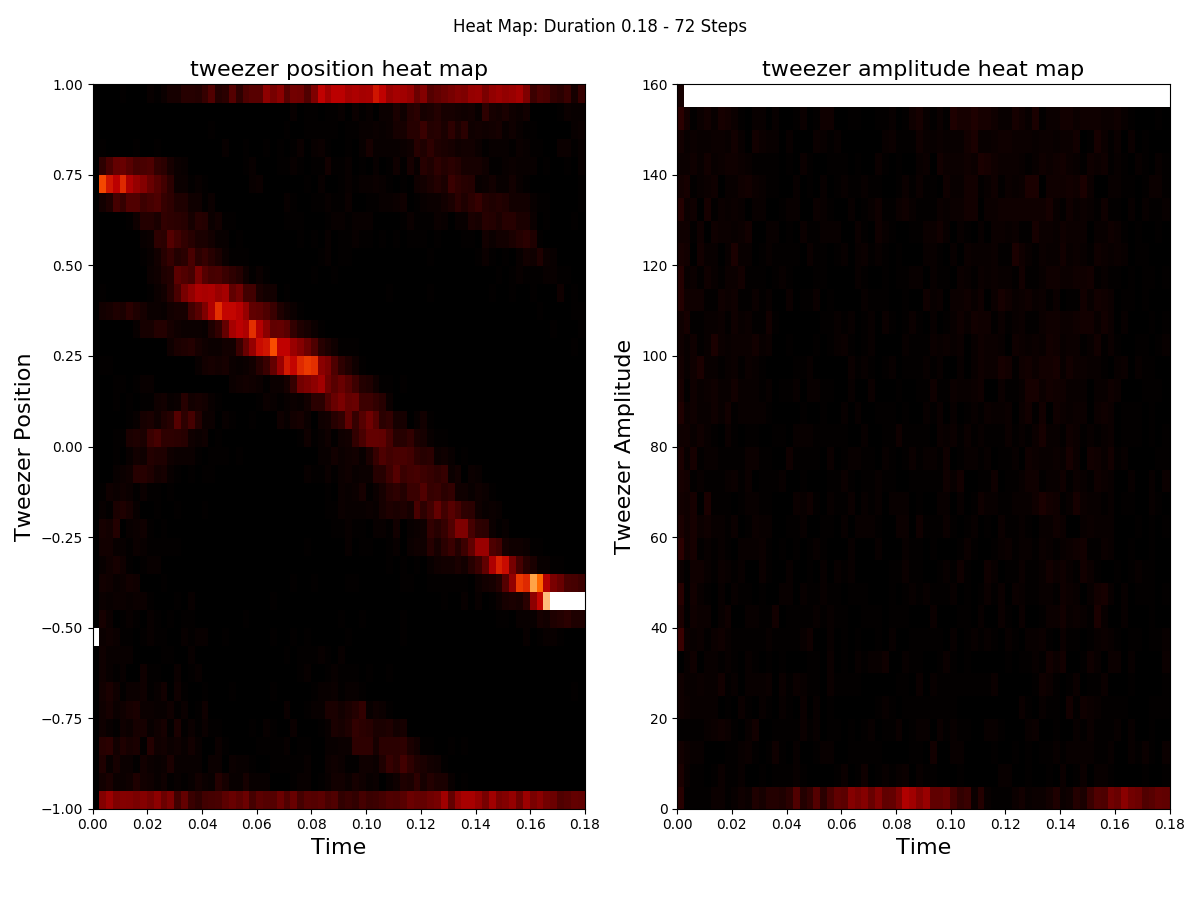}
    \caption{Grape Position and Amplitude Heat Map for duration 0.18, using 72 steps}
  \end{subfigure} 
  \caption{Heat Map Visualization of positions and amplitude for the 200 out of 2000 GRAPE random starts with highest fidelity.}
\label{fig:heat_map}
\end{figure}

 \subsubsection{GRAPE with KASS Seeds}
 \begin{figure}[ht]
\centering \includegraphics[width = 0.7\columnwidth]{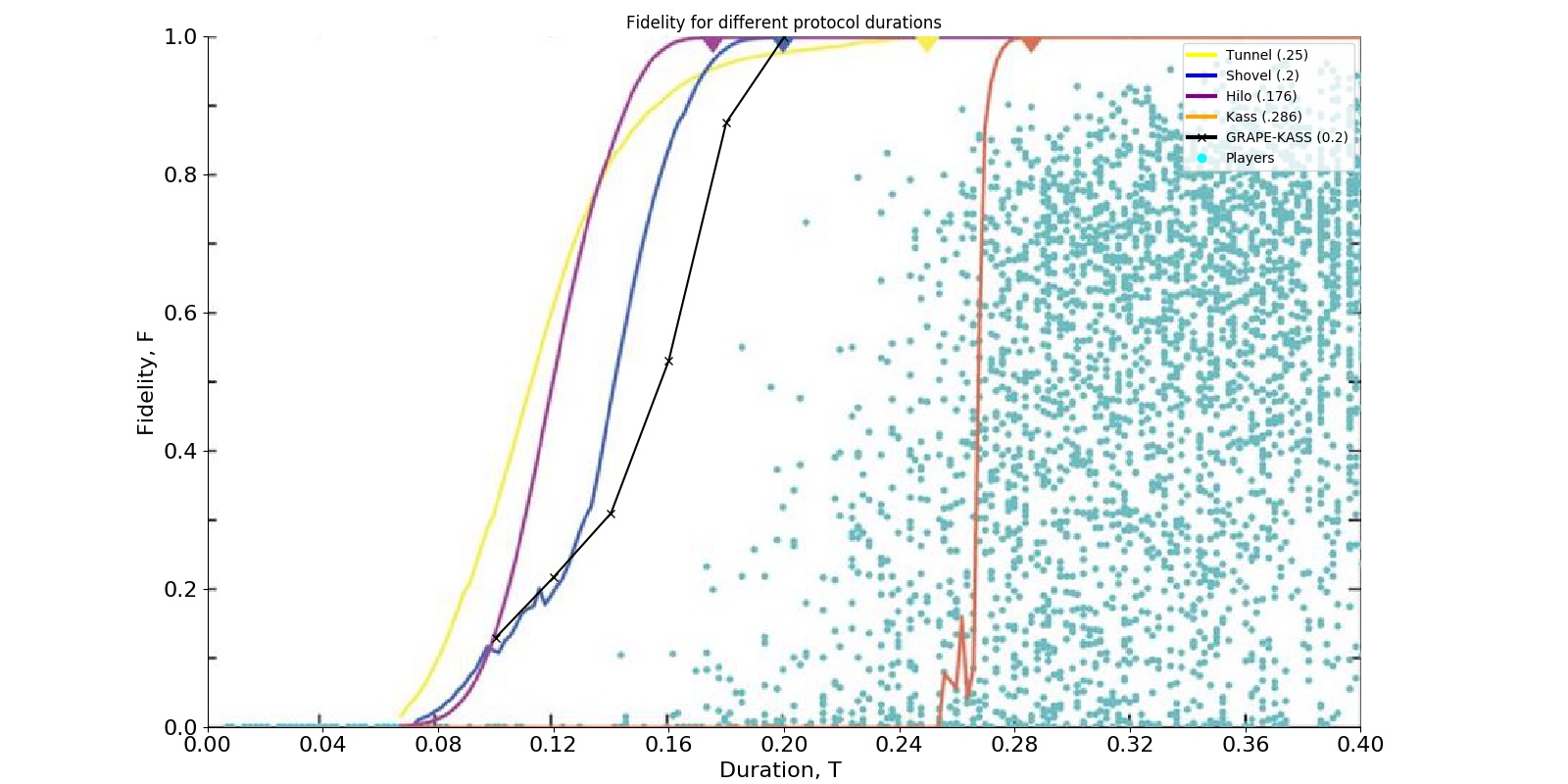}
\caption{Comparing GRAPE with ``KASS'' seeds with the results from \cite{sorensen2016}.}
  \label{fig:kass}
\end{figure}
 It is not hard to make local search algorithms like gradient ascent  perform poorly if you provide bad initial starting points. If you happen to be in this regime, and cannot come up with any better initialization, one could argue that the the obvious solution is to not another algorithm.
 As we have shown above the most trivial of initialization work well, so it is strange that an algorithm like KASS with more elaborate initial values   does so much worse.
 For this reason we decided to test this KASS algorithm a little more. We already know that applied in \cite{sorensen2016} it is pretty bad algorithm for the problem.  But how does the initial starting position in KASS look like.

We quote from \cite{sorensen2016}:
\begin{quotation} We tried different seeding and optimization strategies, and the most successful was a Krotov algorithm using sinusoidal seed functions and sweeps over the total duration (KASS). KASS applies a sweep to an initial random seed $u_i(t)$ where $u_1(t) = A(t)$  is the tweezer amplitude and $u_2(t) = x(t)X$ is the tweezer position. The initial random seed is:
$$
u_i(k\delta_t) = w_i(k\delta_t) + \sum_{n=1}^{N-1} X[n] \sin \left(\frac{n\pi k}{N}\right)
$$
where $N$ is the number of time slices in the path and $\delta_t = 0.002$ is the time discretization. $w_i$ is the motion at constant speed to the initial position of the atom and back again for $A(T) = - 100$. The amplitudes $X[n]$ were selected with a random sign and a norm decreasing as a function of the frequency.
\end{quotation}

There is no reference to where this construction comes from and as such it seems somewhat arbitrary. Furthermore,  the starting values are not fully specified since $X$ is stated to be decreasing in relation to frequency, which could mean many many things, the range is not even specified. For the amplitudes as far as we understand, there is no constant motion but just the constant 100 and then the sinusoids. In conclusion we cannot test this seed structure and try and reproduce the results. What we test instead is simply to ignore the sinusoids and start from linear motion back and forth using half the time on each path as given by $w_1$ and then uniform random amplitudes in $[40, 160]$. In this way the expected amplitude is 100 as it is for KASS as specified. Adding noise to the positions would only make the experiment look very close the GRAPE experiment already performed. We test these starting values with the GRAPE algorithm and get a QSL of 0.2, just as good as the best result starting from player solutions in \cite{sorensen2016}. The full comparison is shown in Figure \ref{fig:kass} where we can see that we get much much better results than the KASS algorithm and also easily beat all the players. Albeit it gets the same QSL of 0.20 as GRAPE with uniform initialization, it deteriorates much faster for lower durations.

\begin{figure}
  \begin{subfigure}[t]{0.5\textwidth}
    \centering
    \includegraphics[width=.4\linewidth]{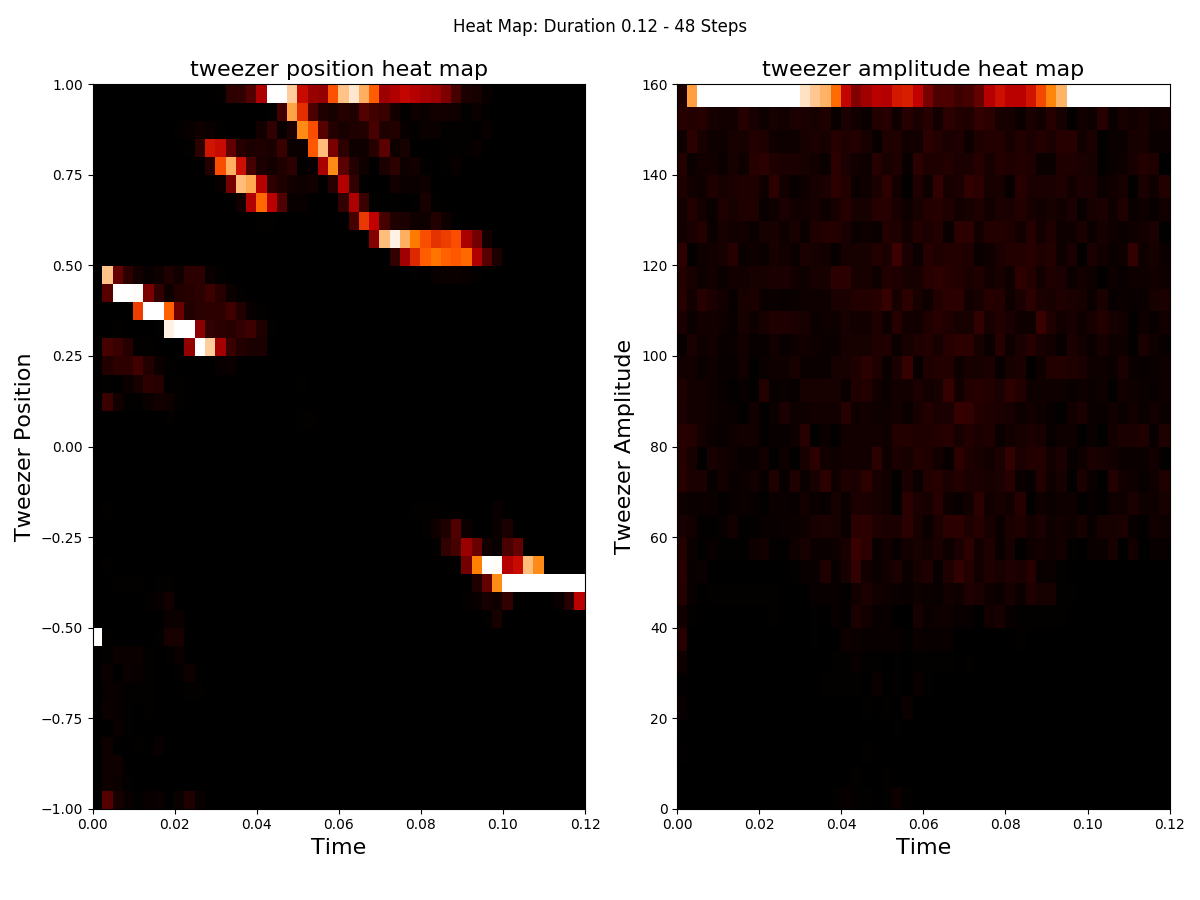}
    \caption{Grape Position and Amplitude Heat Map for duration 0.12, using N=48 steps}
  \end{subfigure}%
  \begin{subfigure}[t]{0.5\textwidth}
    \centering
    \includegraphics[width=.4\linewidth]{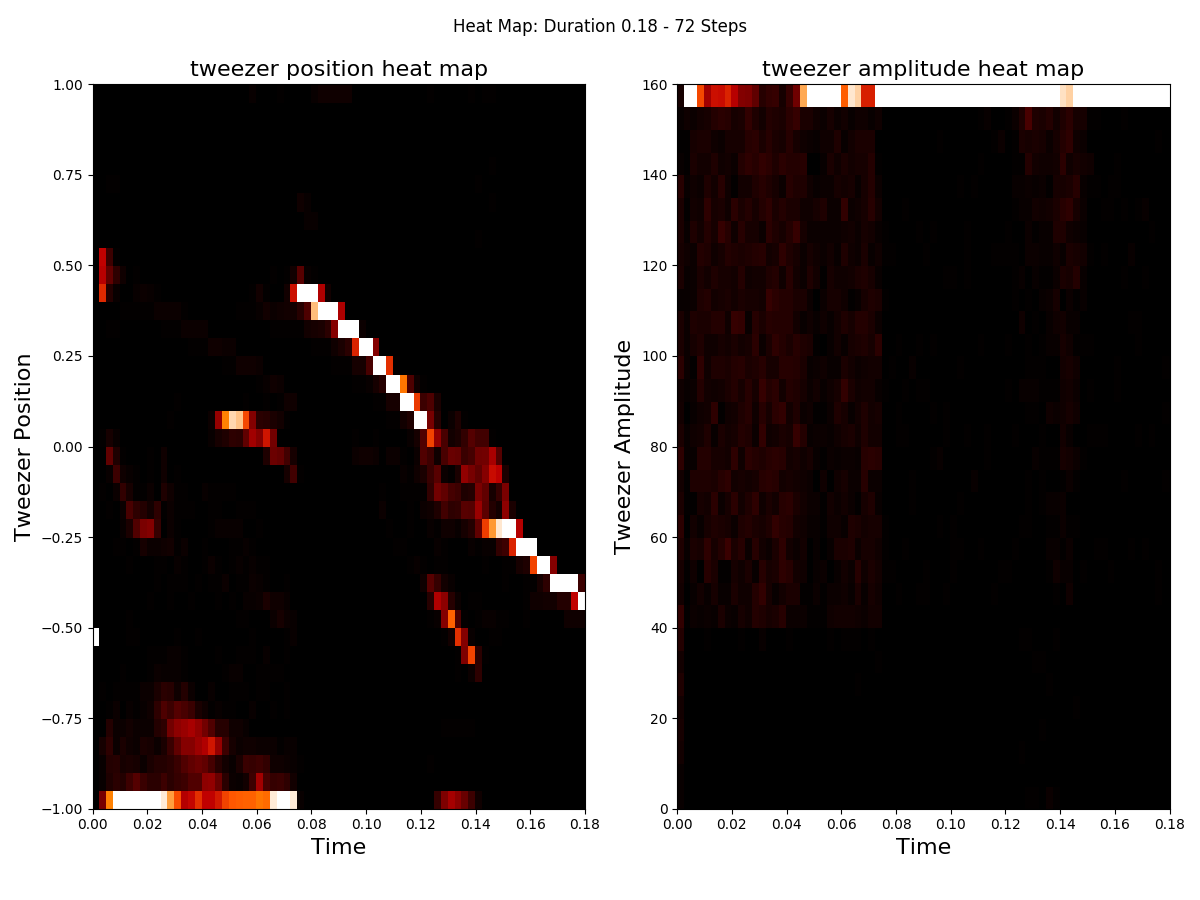}
    \caption{Grape Position and Amplitude Heat Map for duration 0.18, using N=72 steps}
  \end{subfigure} 
  \caption{Heat Map Visualization of positions and amplitude for the 200 out of 2000 GRAPE-KASS protocols with highest fidelity.}
\label{fig:heat_map_kass}
\end{figure}

We have shown two heat maps from GRAPE-KASS  in Figure \ref{fig:heat_map_kass}. Here we can see that for high durations (0.18) the algorithm simply does not know how to transform the initial slow moving of the controllable consistently for instance to moving the tweezer fast to the atom as GRAPE and Stochastic Ascent found. Instead GRAPE-KASS often essentially removes the influence of these steps by moving the tweezer as far away from the atom as possible leaving the atom as it is. For the other half of moving the atom home, GRAPE-KASS  has figured that our very well, including the shaking, and making the amplitude as large as possible. For the low duration the GRAPE-KASS algorithm does seem learn to move to the atom fast, but at the cost of moving the tweezer to the boundary in middle part of the protocol and not maximizing the amplitude in this part, again not using this interval of the protocol for controlling the atom very well. In conclusion this specific way of initializing the GRAPE algorithm, particularly the slow moving of the tweezer towards the atom seems to hinder the algorithm from finding high fidelity solutions for small durations and is generally worse than just uniform random sampling.

\section{The Krotov Algorithm}
\label{sec:krotov}
\begin{figure}[ht]
\centering \includegraphics[width = 0.7\columnwidth]{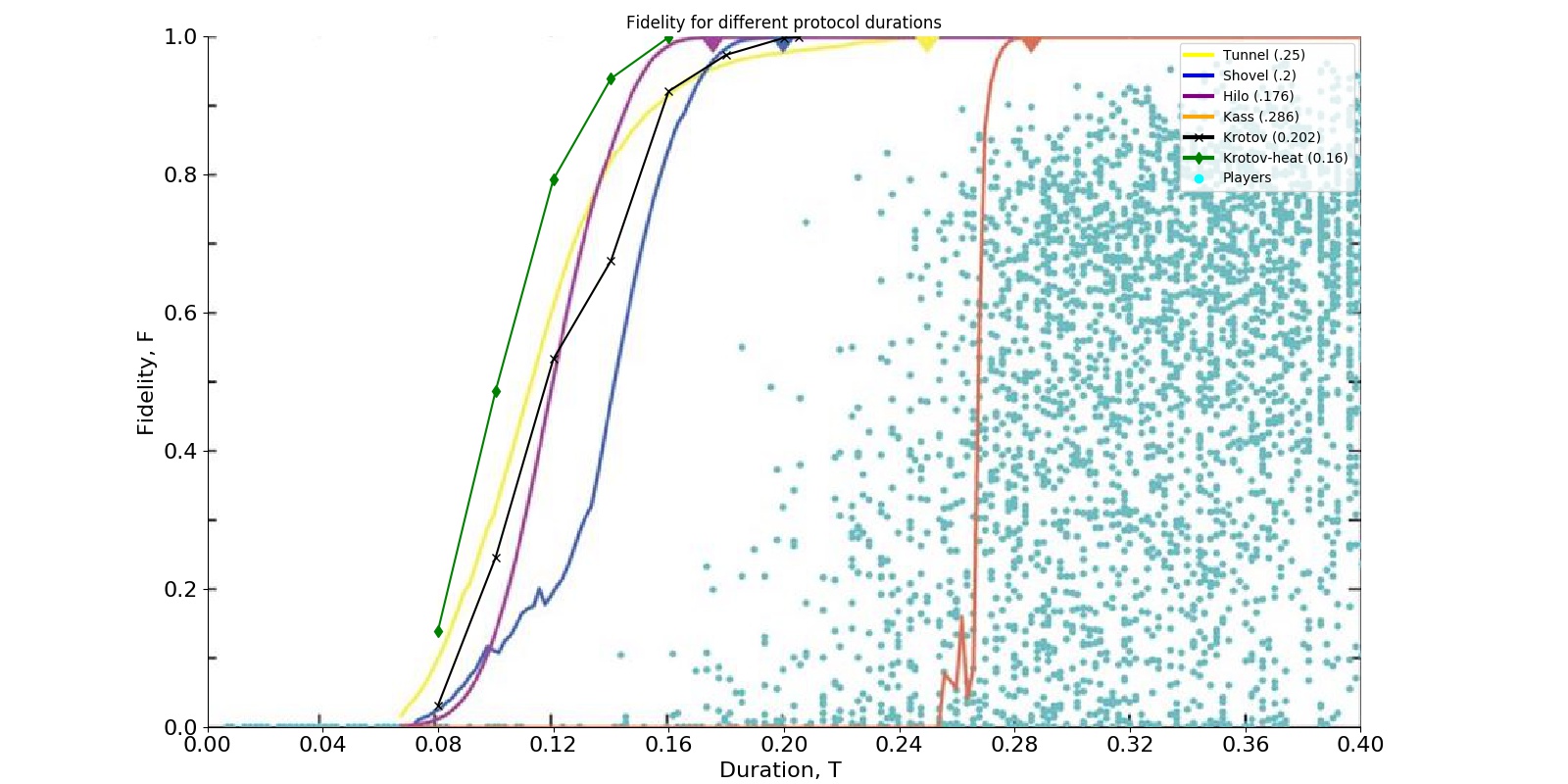}
\caption{Comparing basic Krotov with the results from \cite{sorensen2016}.}
\label{fig:krotov}
\end{figure}

\begin{figure}
  \begin{subfigure}[t]{0.5\textwidth}\centering
    \includegraphics[width=.8\columnwidth]{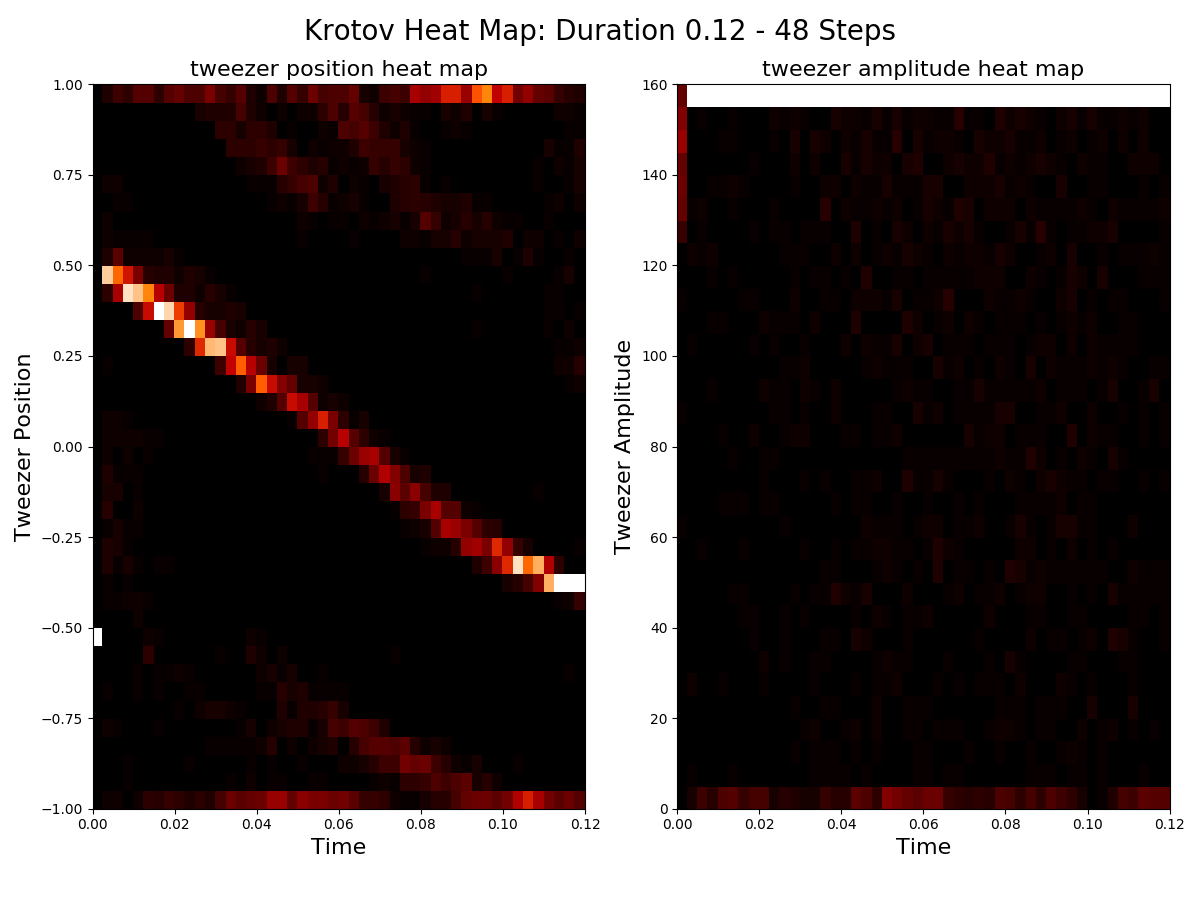}
    \caption{Krotov Position and Amplitude Heat Map for duration 0.12, using N=48 steps}
  \end{subfigure}%
  \begin{subfigure}[t]{0.5\textwidth}\centering
    \includegraphics[width=.8\columnwidth]{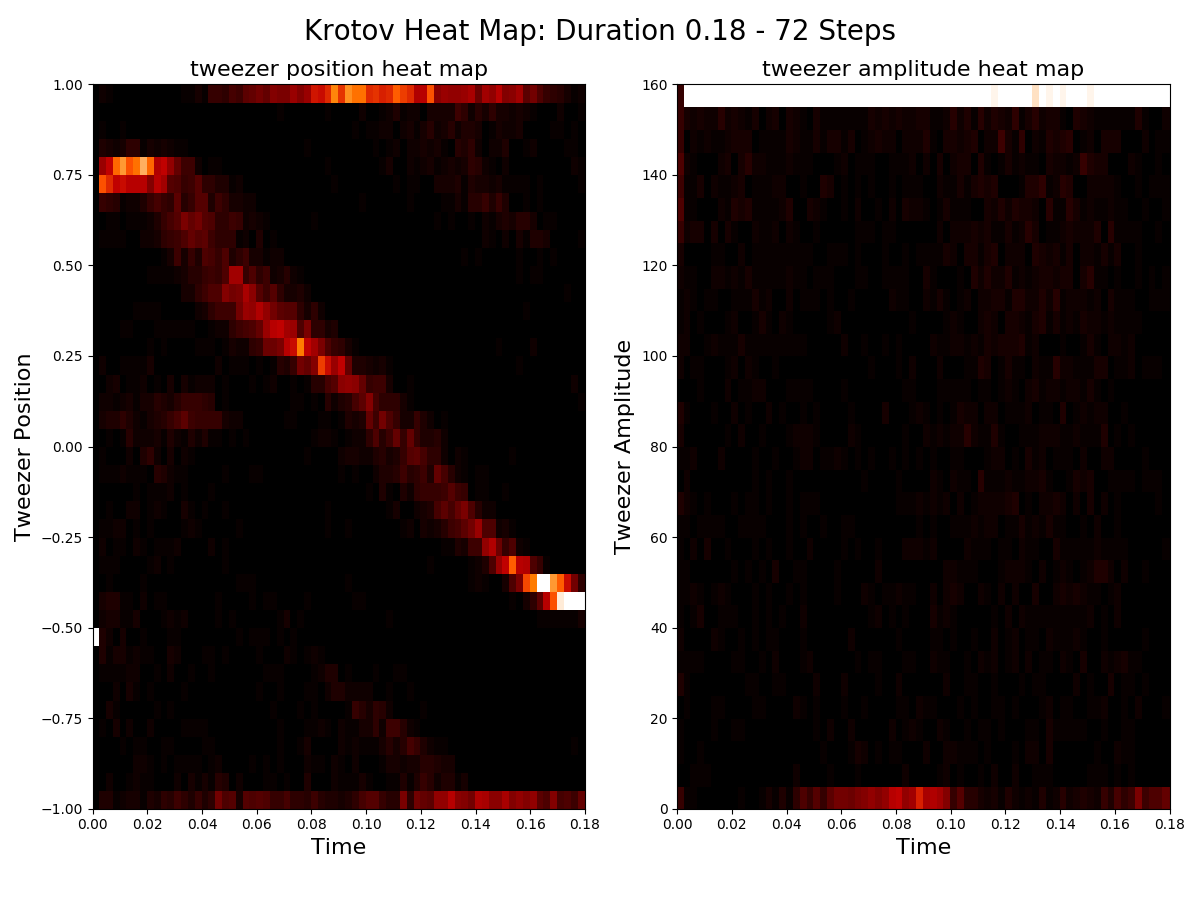}
    \caption{Krotov Position and Amplitude Heat Map for duration 0.18, using N=72 steps}
  \end{subfigure} 
  \caption{Heat Map Visualization of positions and amplitude for the 100 out of 1000 Krotov protocols with highest fidelity.}
  \label{fig:heat_map_krotov}
\end{figure}

In this Section we show the results for doing the same experimental setup we did for GRAPE using a standard Krotov implementation. As for GRAPE we initialize with completely uniform random positions and amplitudes and test the algorithm for different durations. Compared to the KASS algorithm this means we have removed the specific initialization used and the sweeps. A sweep just means the algorithm starts from high durations that are easy, find a good protocol for this duration and then down-sample and run on a slightly smaller duration and then repeating. The results are shown in Figure \ref{fig:krotov}. As we can see from the figure, our simple Krotov implementation, without any help, easily outperforms all players and the KASS algorithm by a large margin obtaining almost the same result as we did with GRAPE.
We have shown two heat maps from the Krotov runs in Figure \ref{fig:heat_map_krotov} which reveal the same as the the heat maps for the GRAPE algorithm did, and as we can see from Figure \ref{fig:krotov} the heat map solution based on Krotov runs is as good as it was for GRAPE matching the results of stochastic ascent.

\section{Experimentally Verifying Theoretical Derived Protocol}
\label{sec:cd_theory}
\begin{figure}[ht]
\centering \includegraphics[width = 0.7\columnwidth]{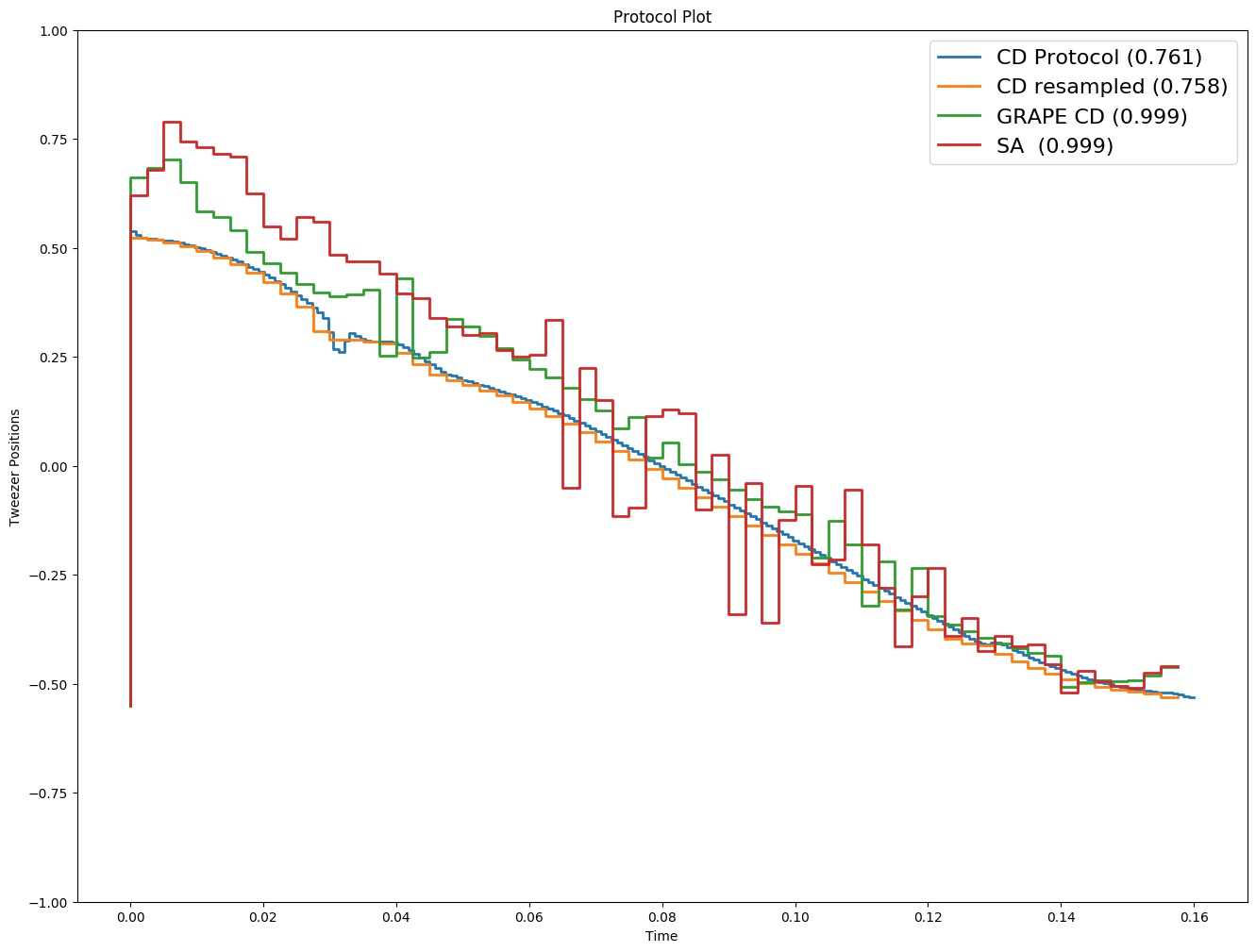}
\caption{Analyzing the theoretically derived counterdiabatic protocol. Number in parentheses is the fidelity achieved for the protocol for duration 0.16.
  CD is the counterdiabatic protocol as derived by D. Sels.
  CD resmapled to 64 steps and fixed first step to -0.55.
  SA is stochastic Ascent with first step fixed to -0.55.
  GRAPE-CD is the result of initializing GRAPE with CD resampled.
  Only CD does not have the initial  position fixed to -0.55}
  \label{fig:cd_prot}
\end{figure}

As we have noted earlier, there is an insignificant difference between the work of Sels \cite{Dries, DriesarXiv} and the work in \cite{sorensen2016}
due to the first step of the protocols discussed by D. Sels. In \cite{sorensen2016} the controllable tweezer is fixed at the starting position in the first step, while in \cite{Dries, DriesarXiv} this positions is optimized by the method and placed very close to the atom instead, as keeping the first step fixed at $-0.55$ just wastes time.  As we have amply shown, for Stochastic Ascent, this makes no 
real difference at all. In order to remove any doubt that the analytical protocols derived by Sels \cite{Dries} using counterdiabatic driving (CD) indeed
does capture the whole problem, we show in Figure \ref{fig:cd_prot} a comparison between the analytical protocols, the  
results using stochastic ascent (SA), and GRAPE initialized with the CD protocol. To ensure that there can be no issues for the comparison, we have resampled the CD protocol and fixed the starting point to -0.55. As we can see this does not really change the fidelity achieved by the protocol, just like it was the case for stochastic ascent. Indeed, we see in Figure \ref{fig:cd_prot} that there is a near-perfect match between the raw CD protocol and the resampled version and both match the fidelity reported for the protocol by D. Sels for the given time duration \cite{Dries, DriesarXiv}.
It is also clear from the figure that the GRAPE algorithm initialized with this counterdiabatic protocol clearly beats all algorithm from \cite{sorensen2016} obtaining a QSL of 0.16.

We note that if we initialize the tweezer positions for GRAPE with the solution for a single tweezer version of the problem analyzed in \cite{Dries} we get a QSL 0.16 matching the best results achieved. As described in \cite{Dries} the simple line gives the smallest magnitude of the acceleration, which we  need to avoid to avoid spilling. However, the line has the issue that the velocity does not cancel out at the beginning and the end. This  is handled  by using a third order polynomial instead in \cite{Dries}. However, a gradient ascent algorithm may be able to take care of both these issues, and as the results shows it does. Starting with this simple initialization for gradient ascent by picking points on a line from end position to starting position gets a better QSL than the best algorithms from \cite{sorensen2016}, actually obtaining the same QSL  as the stochastic ascent algorithm and the Krotov and GRAPE heat map algorithms, in one try.

\begin{figure}[t]
\centering \includegraphics[width = 0.7\columnwidth]{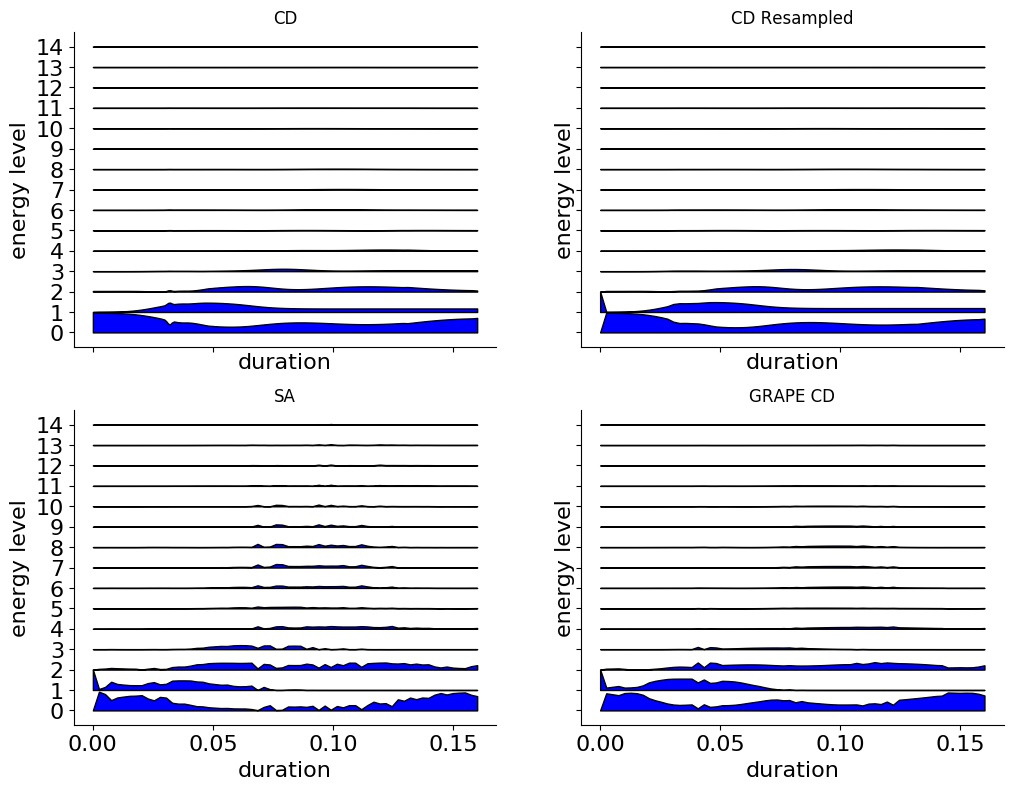}
\caption{Excitation plots for the 15 lowest energy levels for the protocols in Figure \ref{fig:cd_prot} in particular the theoretically derived counterdiabatic protocol by D. Sels \cite{Dries}}  
\label{fig:exitation}
\end{figure}

In the work \cite{sorensen2016}, there is also a discussion of the excitation of the atom during the protocols (see Figure 3c of \cite{sorensen2016}). In 
Figure \ref{fig:exitation}, we show the excitation for the same experiments as in Figure \ref{fig:cd_prot}. The analytical protocol derived by 
D. Sels uses counterdiabatic driving and moves to an accelerated frame to solve the problem. In the new frame Sels
looks for the ground state. However, once this is transformed to the original frame, the atom does not stay in the 
ground state. This can be seen in Figure \ref{fig:exitation} very clearly. Here we see that the CD protocols are very similar to the 
numerical protocols with SA and GRAPE. This confirms once again that the theoretically derived CD protocol discussed 
in \cite{Dries} has correctly captured the problem. The difference between CD protocol and the numerical protocols, is due to 
the shaking motions that are seen in Figure \ref{fig:cd_prot} as well. However, the overall profiles are consistent. We may again 
conclude that the work in \cite{Dries, DriesarXiv}, both theoretical protocols and numerical experiments, are correct, capture the whole 
problem, and show without a doubt that theoretical and numerical algorithms far outperform human players in the BringHomeWater game.

We have also tests a theoretically derived counterdiabatic protocol for duration 0.22. The analysis of the protocol for this duration shows the exact same thing and we leave it out.

\section{Final Comments}
\label{sec:conclusion}
We have experimentally verified the results of the simple Stochastic Ascent algorithm from \cite{Dries}. We have tested different possibilities and shown that this does not change the performance of the simple Stochastic Ascent algorithm.  We also experimentally tested the theoretically derived CD protocol from \cite{Dries}. We conclude that the results, discussion and conclusions in \cite{Dries} are verified. Furthermore, we have shown that a basic gradient ascent and a basic Krotov algorithm also easily outperform all players and the KASS algorithm. We note that the QSL of 0.16 achieved by different methods in this note, in particular stochastic ascent and gradient ascent and Krotov initialized properly may be improved for instance by testing slightly smaller durations.

As pointed out by D. Sels in \cite{Dries}, the player solutions are variations of the intuitive solution that move the tweezer to the atom  and then slowly back again with a few shakes to avoid spilling. Using such protocols (while ignoring the shakes) gives essentially the initialization that seems to be the main result achieved in \cite{sorensen2016}, albeit they use a different algorithm based on Krotovs method, that is worse than standard gradient ascent and basic Krotov, on top of it. 
Thousands and thousands of gamers have played the game but it seems clear that they have not come up with anything that, even when combined with the best algorithms by  \cite{sorensen2016}, is as good as the basic stochastic ascent algorithm or the basic gradient ascent or krotov algorithm initialized from a simple heat map analysis of several GRAPE or Krotov runs. The algorithms also reveal two similar but different strategies for solving the problem; one for short durations and one for larger. 
We note that we  only tested basic algorithms from mathematical optimization and quantum control, and that there are many other more elaborate algorithms designed for hard computational problems that almost certainly also outperform both gamers and the player inspired algorithms from \cite{sorensen2016} as well. To summarize, all the conclusions made in \cite{sorensen2016} regarding algorithms and optimization is entirely based on poor application of algorithms.  From the results it is clear that human players and human intuition with the help provided by the game designers is clearly outperformed by theory and algorithms. This suggests that this will also be the case for more complicated problems like for instance the superposition version of BringHomeWater as introduced in \cite{Dries}.
KASS was claimed to best numerical algorithm  for Bring Home Water by \cite{sorensen2016}, which seemingly lead to the conclusion that gamers beat algorithms. As shown by D. Sels in \cite{Dries} and verified in this note, this is clearly not true, gamers do not even come close the performance of standard algorithms. In fact the KASS algorithm is exceptionally poor when compared with standard implementation of gradient ascent, and the Krotov algorithm which KASS is based on. In conclusion it is clear that that KASS implementation is a very poor algorithm for the problem. Thus the fact that players are able to beat KASS on a few occasions has no significance. Stated differently, the conclusion from \cite{sorensen2016} that ``\emph{Players succeed where purely numerical optimization fails} is false. 


The other main result from \cite{sorensen2016} that ``\emph{Using player strategies, we have thus developed a few-parameter heuristic optimization method that efficiently outperforms the most prominent established numerical methods}''. As argued, this can only be true if you define the \emph{most prominent established numerical algorithms} as KASS and CRAB, and as such is completely misleading.
As shown by D. Sels in \cite{Dries} and underlined by the results in this paper, this is clearly false as well. In particular, stochastic ascent is simpler, it is more efficient, it does not require hundreds of trials to get the a good solution and it achieves  better results. The same holds true for GRAPE and Krotov. While the best method in \cite{sorensen2016} is better than uniform initialization GRAPE, a reasonably initialized GRAPE and Krotov easily beats it, and a good initialization can be made from theory or a standard analysis of protocols extracted from a few runs of the algorithm taking a few hours of computer time on a standard work station.

On a side note the results in \cite{Dries, DriesarXiv} and this note  also contradict the argument from \cite{sorensen2016}, that  ``\emph{The choice of an initial seed is a central challenge in such complex optimizations}'' as BringHomeWater and the statement that \emph{We stress that any optimal control effort requires a substantial amount of optimizations of different seeds to find good solutions.}, since no fancy initialization, using human players or otherwise, is needed.


\appendix 

\section{Stochastic Ascent Algorithm}\label{appa}
The Hamiltonian's defining the system in BringHomeWater is defined as follows.
Let $B$ be the amplitude of the fixed tweezer,  then the initial state of the atom is the ground state of the Hamiltonian 
$$
H = p^2/2m - B e^{-(x-x_\textrm{start})^2/(2\sigma) } 
$$
and the final state is the ground state of the Hamiltonian
$$
H=p^2/2m- B e^{-(x-x_\textrm{end})^2/(2\sigma)} 
$$
Let $A_t$ be the amplitude of the controllable tweezer, and $x_t$ be the position of the controllable tweezer at time $t$ then the time evolution operator Hamiltonian is
\begin{equation}
  \label{eq:schr}
H_{t} = p^2/2m - A_t e^{-(x-x_t)^2/2\sigma} - B e^{(x-x_\textrm{start})^2/2\sigma} 
\end{equation}

Following \cite{Dries}, we implement a Stochastic Coordinate Ascent algorithm for BringHomeWater as shown in Figure \ref{fig:cascent}.
As in \cite{Dries}, the algorithm does not optimize over the amplitude of the controllable tweezer, keeping the amplitude fixed at the maximum allowed value all the time as discussed earlier.
Secondly, the algorithm uses a reasonably small  discrete set of positions for the controllable tweezer which is simply  a uniformly 
spaced grid of some fixed size $s$. Note that the same grid is used for all steps of the protocol. This allows computing all the needed unitary matrices only once before the iteration starts.

Now consider the configuration space of the problem. Let \emph{s} denote the number of possible positions for the controllable tweezer.
In any protocol we have to specify a number of steps that we can take, denote this by $N$. The configuration space is therefore $s^N$. For any reasonable 
values of $s$ and $N$, $s^N$ will be an astronomical number, and testing all  protocols is not feasible.

\subsection{Numerical Computations}
\label{app:numerical}
To compute the fidelity of a protocol, we use a grid of $h_g$ of uniformly spaced points to discretize the Schrodinger equation and Finite Difference approximation for the Laplacian.
This requires computing matrices for the three parts in the Schr{\"o}dinger Equation (Equation \ref{eq:schr}).
First, define a uniformly spaced grid $g_h$ over $[-1, 1]$ with $h_g$ uniformly separated points (including -1, +1) at distance $d_h = 2/(h_g-1)$.
For the Laplacian, generate a finite difference matrix as $\frac{0.5}{d_h^2}$ times a tridiagonal matrix with two on diagonal and -1 above and below the diagonal.
\begin{equation*}
\fdm = \frac{0.5}{d_h^2}\begin{bmatrix}
 2 & -1     & 0      &       & 0     & 0  \\
-1  & \ddots & \ddots & \ddots &    \    & 0 \\
0 & \ddots & \ddots & \ddots & \ddots &    \\
 & \ddots & \ddots & \ddots & \ddots & 0  \\
0  &        & \ddots & \ddots & \ddots & -1 \\
0 & 0      &        & 0     & -1      & 2
\end{bmatrix}
\end{equation*}

For the part of the Schrodinger Equation for the fixed tweezer, $B e^{-(x-x_\textrm{start})^2/2\sigma}$, generate a diagonal matrix $M_B$ where the ith entry is $B e^{-(h_g[i] - x_\textrm{start})^2/(2\sigma)}$
In vectorized notation this can be written as 
$$
\textrm{diag}(\exp(-((g_h - x_\textrm{start})^2) / (2\sigma^2))
$$
where $g_h$ is the list of grid points used for discretization.

For the final part of the equation for the controllable tweezer, $A e^{-(x-x_\textrm{tweezer})^2/2\sigma}$, 
compute a matrix for each allowed position of the tweezer and for each such position, construct a matrix as for the fixed tweezer.
Thus, for each tweezer position $t_p$, make a diagonal matrix where the $i$'th entry is $A e^{-(t_p - g_h[i])^2/(2\sigma)}$.
For position $t_p$ this can be written in vectorized notation as 
$$
\textrm{diag} (A \exp(- (g_h-t_p)^2/(2\sigma^2)))
$$
where again $g_h$ is the list of grid points used for discretization.

Finally, for every potential tweezer position $t_p$ with associated matrix $M_p$ compute the Hamiltonian matrix
$$
h_p = \fdm - M_b - M_p
$$
, and to get the unitaries compute
\begin{equation*}
  \label{eq:unitaries}
U_p = \exp(h_p \cdot (T/N) (-i)) .
\end{equation*}
This defines all the required unitary time evolution matrices, and for a given protocol the time evolution operator is the a product of an the corresponding multi-set of these ordered by the protocol.
For the initial and target state compute the eigenvector with the smallest eigenvalue of the appropriate corresponding Hamiltonian matrix. In particular the starting state $\ket \psi$ is the eigenvector corresponding to the smallest eigenvalue of $\fdm - M_b$.
For the target state, create the matrix $\bar{M_b}$ symmetric to $M_b$ just for $x_\textrm{end}$ instead of $x_\textrm{start}$, i.e. a diagonal matrix with the $i$'th entry equal to  $B e^{-(h_g[i] - x_\textrm{end})^2/(2\sigma)}$.
Now $\bra \phi$ is the eigenvector corresponding to the smallest eigenvalue of the matrix $\fdm - \bar{M_b}$. Notice that the controllable tweezer is turned off for both the start and end state.

\subsection{Implementing Algorithm}
\label{app:sa_implementation}
With the math behind us the the Stochastic Ascent algorithm can now be implemented  with a fairly short python program using numpy and scipy
(or any other language if preferred). We always use $m=1, B=130, L=1.1, \sigma = \frac{1}{8}$ as defined in \cite{Dries}. The remaining parameters depends on the experiment.
A description of the algorithm is also contained in \cite{Dries} but for completeness we include (a shorter) one here in Figure \ref{fig:saa}.

\begin{figure}[ht]
\begin{tabular}{l}
  \textbf{Stochastic Ascent Algorithm}\\
  \hline
  Input Parameters $A, B, \sigma, T, N, s$\\
  \hline \\
1. Compute target state, $\bra \phi$, the initial state $\ket \psi$, and unitary matrices $U_0,\dots, U_{s-1}$ (Appendix \ref{app:numerical}).\\
2. Randomly initalize a protocol of $N$ random integers $[x_N,\dots,x_1]\in \{0,1,\dots,s-1\}^N$.\\
3. Let $I$ be random permutation of $\{1,\dots,N\}$\\
4. for $i \in I$\\
4a. $\quad$ Compute prefix and suffix of fidelity computation, $a = \bra \phi \prod_{j=i+1}^N U_{x_j}$ and $b = \prod_{j=1}^{i-1}U_{x_i} \ket \psi$.\\
4b. $\quad$ $x_i =  \argmax_{k\in \{0, \dots, s-1\}} | a U_k b |^2$ \\
5. If fidelity does not improve in step 4 then return current protocol, else goto step 3
\end{tabular}
\caption{More detailed Stochastic Ascent Algorithm for BringHomeWater (Quantum Moves) from \cite{Dries}}
\label{fig:saa}
\end{figure}

In $\cite{Dries}$, $s$, the number of positions available for the controllable tweezer is set to 128. With 128 as the number of tweezer positions we cannot even place the controllable tweezer at the atom or at the target position! For the general result of showing that the Stochastic Ascent algorithm easily and clearly beats human gamers 128 is more than enough. If it is important to get fidelity above 0.999 as the goal set in \cite{sorensen2016} it helps to increase this parameter for instance doubling it or setting it to 201 to ensure the controllable tweezer can be placed on $x_\textrm{start}$ and  $x_\textrm{end}$ if needed. This will also of course make the algorithm slower by a constant factor. 
The grid size $h_g$ for discretizing the Hamiltonians is a numerical precision parameter which we set to 512, albeit for fast tests using a smaller value gives almost the same results. 
We checked that the discretization indeed changes very little to the results at that point.
To fix the first position of the controllable tweezer in the protocol, initialize $x_1$ to $-0.55$ (or closest grid point) and then do not optimize that position in the algorithm  by not including it in the random permutation $I$. To allow more amplitudes than  $A=160$, we add the necessary unitary matrices as described above, and for each step optimize over position, amplitude pairs, scaling the number of possibilites to check with the discretization of the amplitudes considered. We have only tested this with a few parameter settings. The results from these experiments indicate that the algorithm naturally slows down but still finds solutions of the same quality, and as the theory suggest using the maximal value for the amplitude works. 

\subsection{Gradient Ascent - GRAPE}
To implement gradient ascent, we implement the fidelity computation, as described above for Stochastic Ascent, in TensorFlow and use automatic differentiation. The number of steps, $N$, for a given duration, $T$, is set as for stochastic ascent such that $T/N = 0.0025$. For each duration $T$ tested we ran several GRAPE invocations from completely uniform random starting points. 
Compared to Stochastic Ascent, GRAPE is more time consuming and in order to save time, we ran the main bulk of experiments using imprecise computation by simply setting a low value for the grid size $h_g$ and running each experiment only a limited amount of iterations. We ran 2000 experiments starting with,  $h_g=32$ and ran it for at most 2000 iterations or until it had stopped improving (gradient to small).  Then the best protocol found like this is considered again, we increase the grid size back by doubling and restarting the iteration, until we hit a grid size of 512. While there is no guarantee that best model for the small $h_g$ is the best model to continue from, it seems a reasonable choice and small tests confirm that this, and saves a lot of computing time.   
Since ADAM \cite{kingma2014adam} is the standard recommended optimizer in TensorFlow (albeit for deep learning) we use that for the experiments and changed only the standard learning rate parameter.
We start with an initial learning rate of 0.1, or 0.01 and reduce it during the algorithm when the fidelity stops improving.
It takes around four to six hours on a standard work station to run all 2000 random starts and optimizing the best found protocol from there takes even less.
For a full test one should really test different optimization methods and very their hyper parameters. Since such a large experimental evaluation is not the point of that note, we leave that to the interested reader. Since we only tested a few things we believe that even better results can be achieved for GRAPE with more experiments and tuning.

To ensure the tweezer starts as in the game we always fix the first of the $N$ position available to $-0.55$ and never update it in the algorithm.

\begin{table}[ht]
  \centering
\begin{tabular}{cc}
  \begin{subfigure}[t]{0.45\textwidth}\centering\includegraphics[width=1\columnwidth]{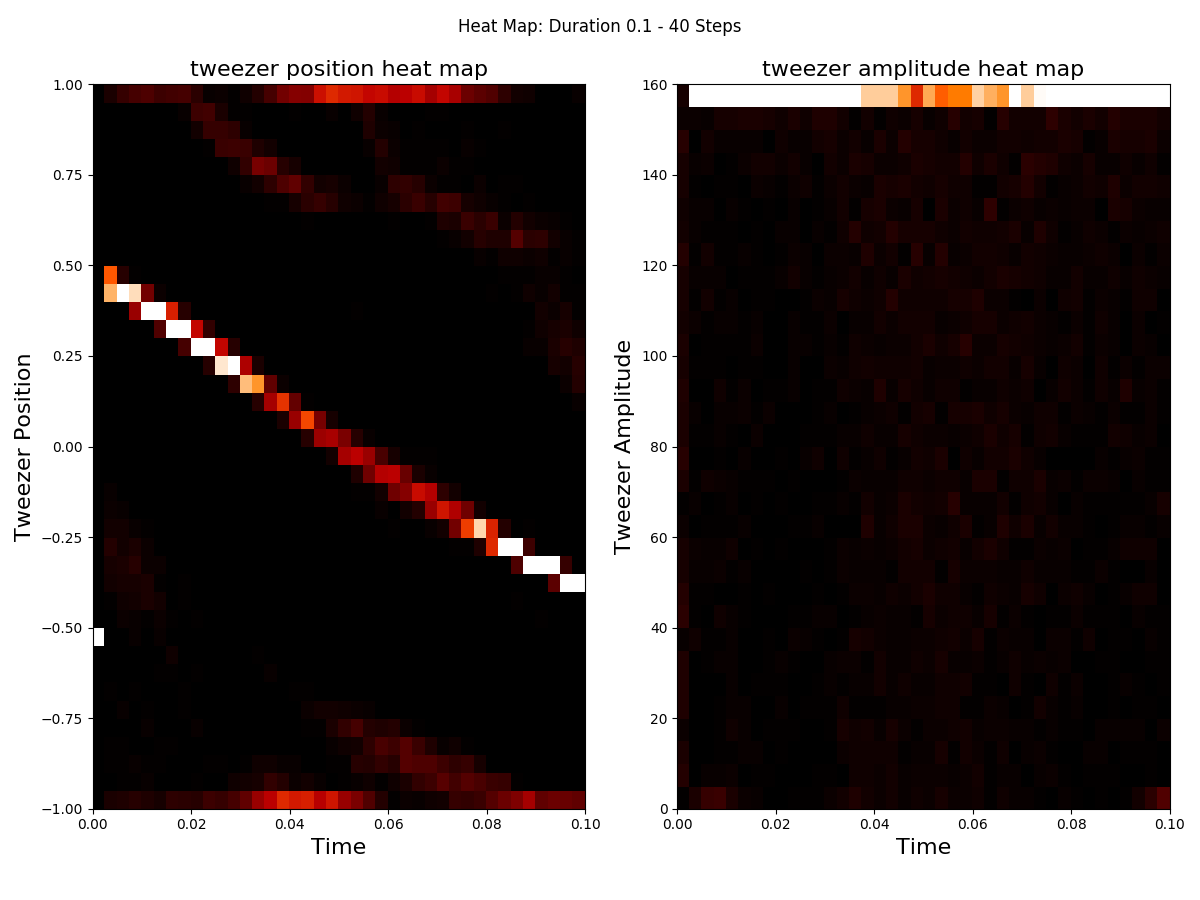}
    \caption{Grape Position and Amplitude Heat Map for duration 0.1, using 40 steps}
    \label{fig:ha}\end{subfigure}
  &
  \begin{subfigure}[t]{0.45\textwidth}\centering\includegraphics[width=1\columnwidth]{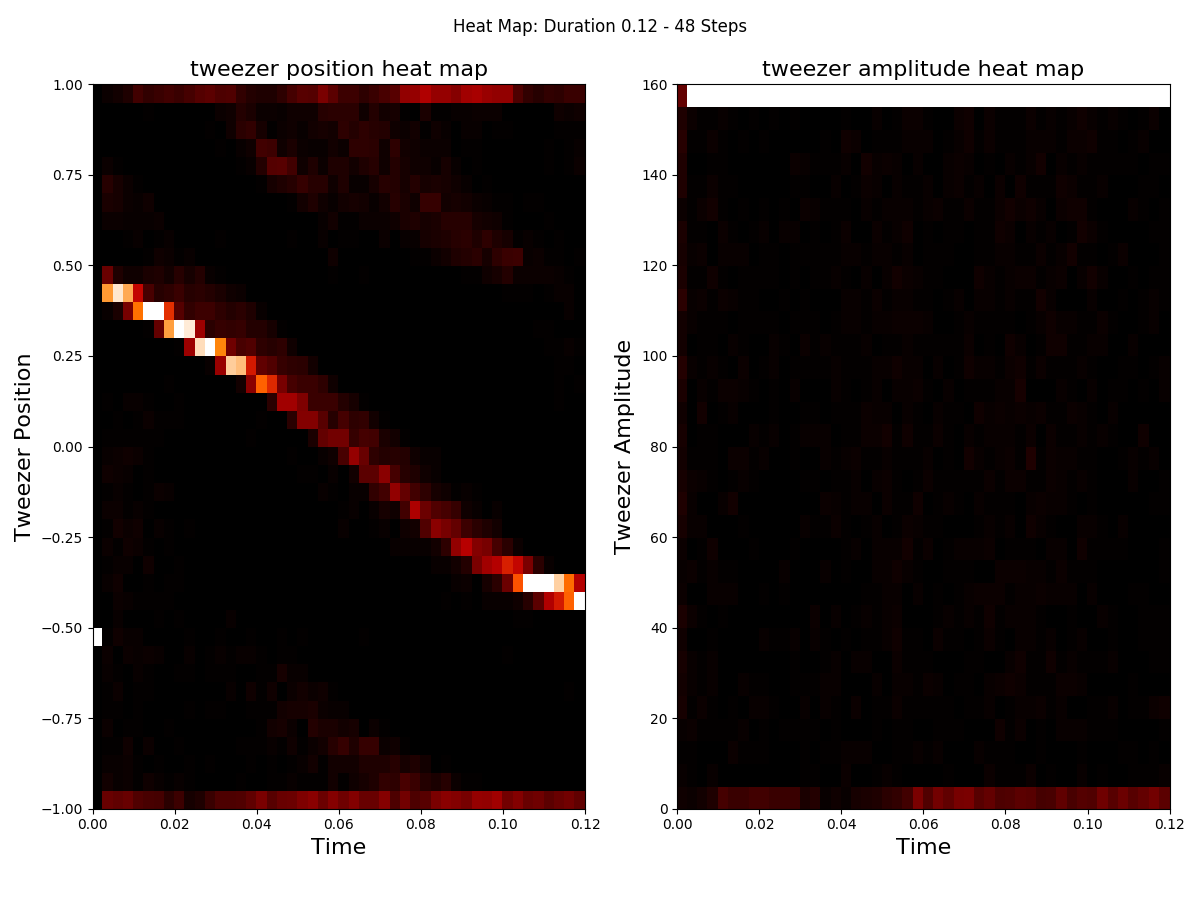}
    \caption{Grape Position and Amplitude Heat Map for duration 0.12, using 48 steps}
    \label{fig:hb}\end{subfigure}\\
\begin{subfigure}[t]{0.45\textwidth}\centering\includegraphics[width=1\columnwidth]{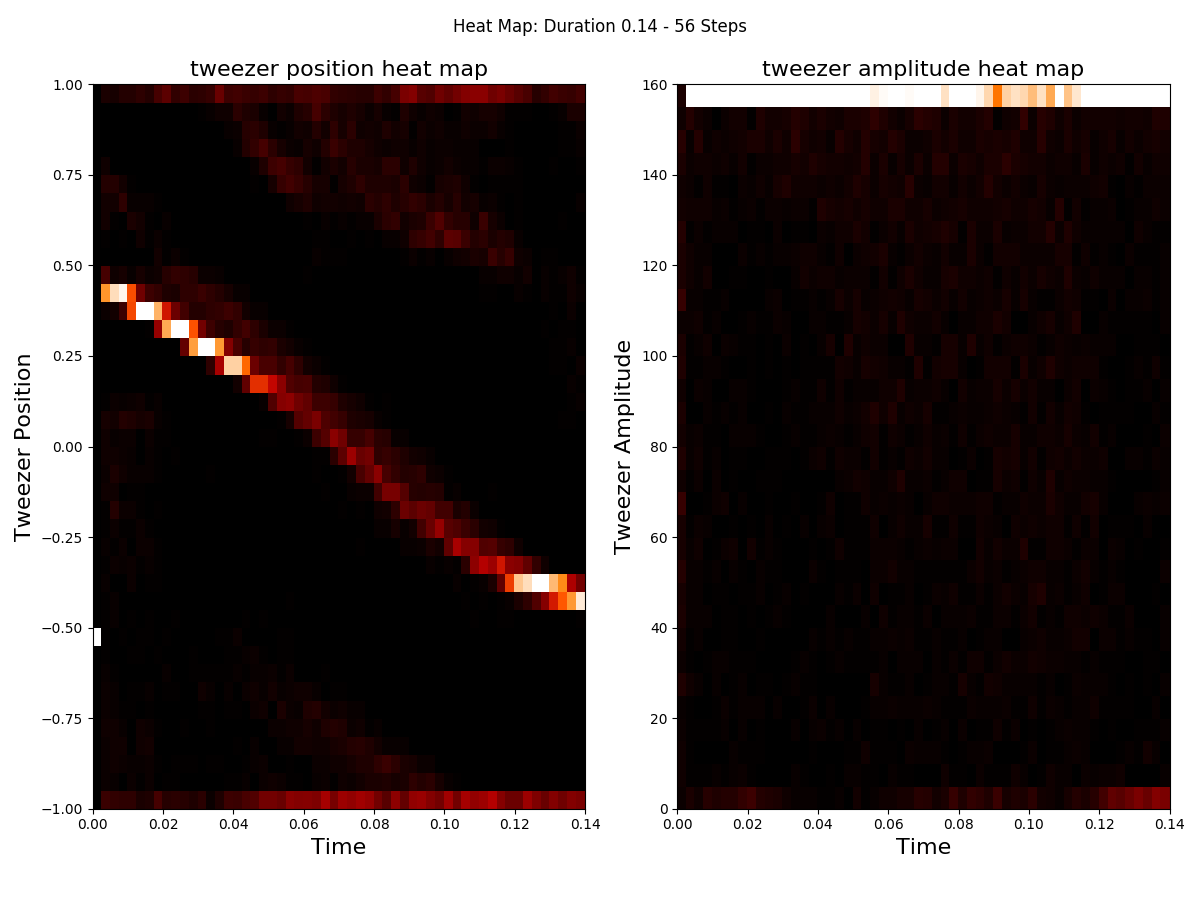}
  \caption{Grape Position and Amplitude Heat Map for duration 0.14, using 56 steps}
  \label{fig:hc}\end{subfigure}
&
\begin{subfigure}[t]{0.45\textwidth}\centering\includegraphics[width=1\columnwidth]{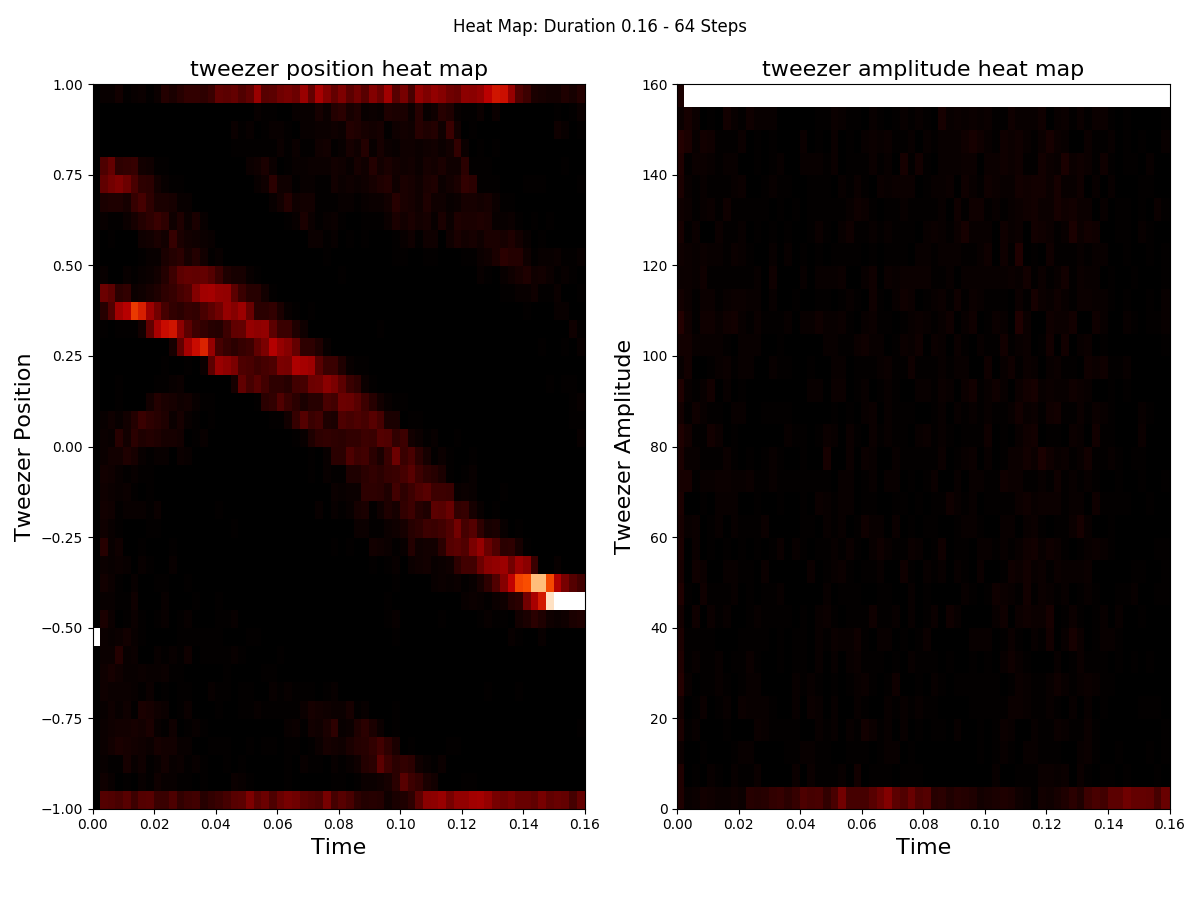}
  \caption{Grape Position and Amplitude Heat Map for duration 0.16, using 64 steps}
  \label{fig:hd}\end{subfigure}\\
\begin{subfigure}[t]{0.45\textwidth}\centering\includegraphics[width=1\columnwidth]{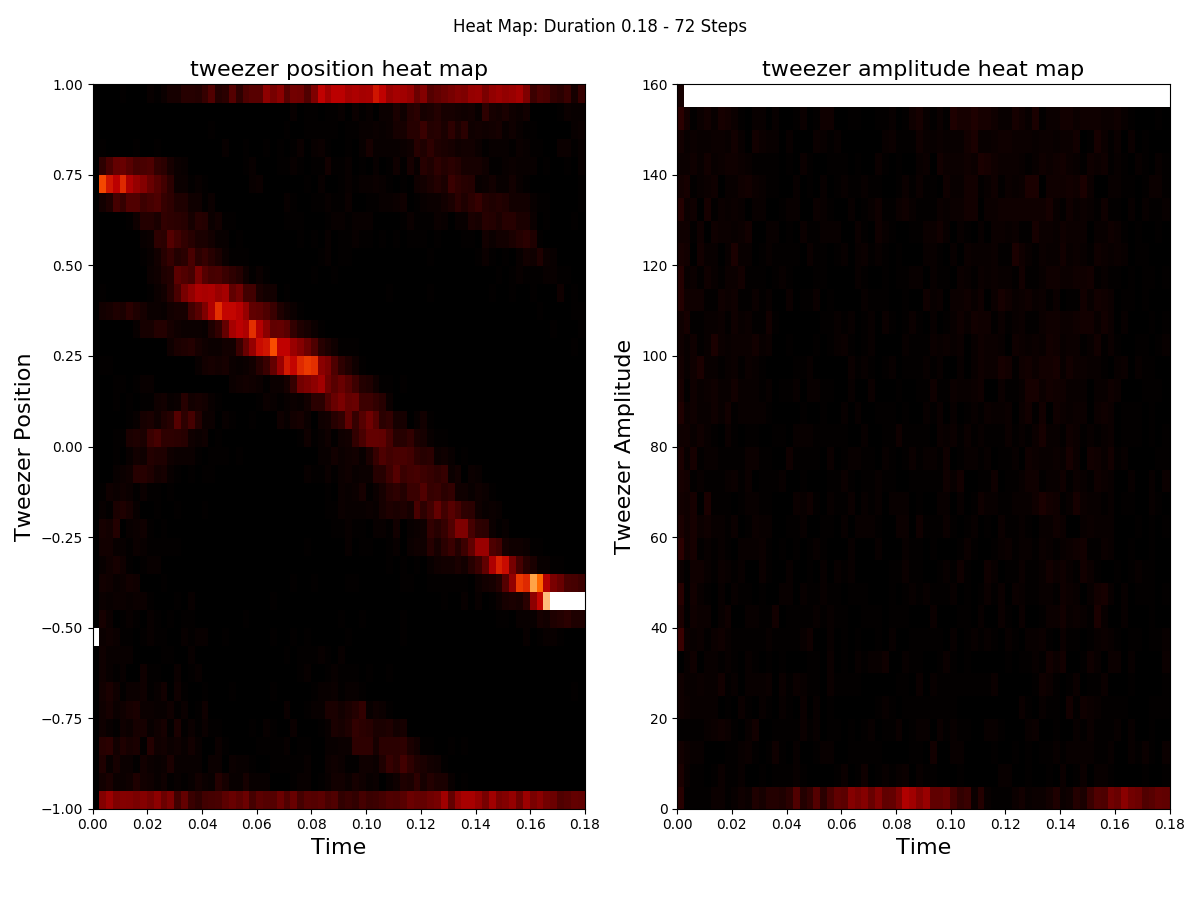}
  \caption{Grape Position and Amplitude Heat Map for duration 0.18, using 72 steps}
\label{fig:he}\end{subfigure}
&
\begin{subfigure}[t]{0.45\textwidth}\centering\includegraphics[width=1\columnwidth]{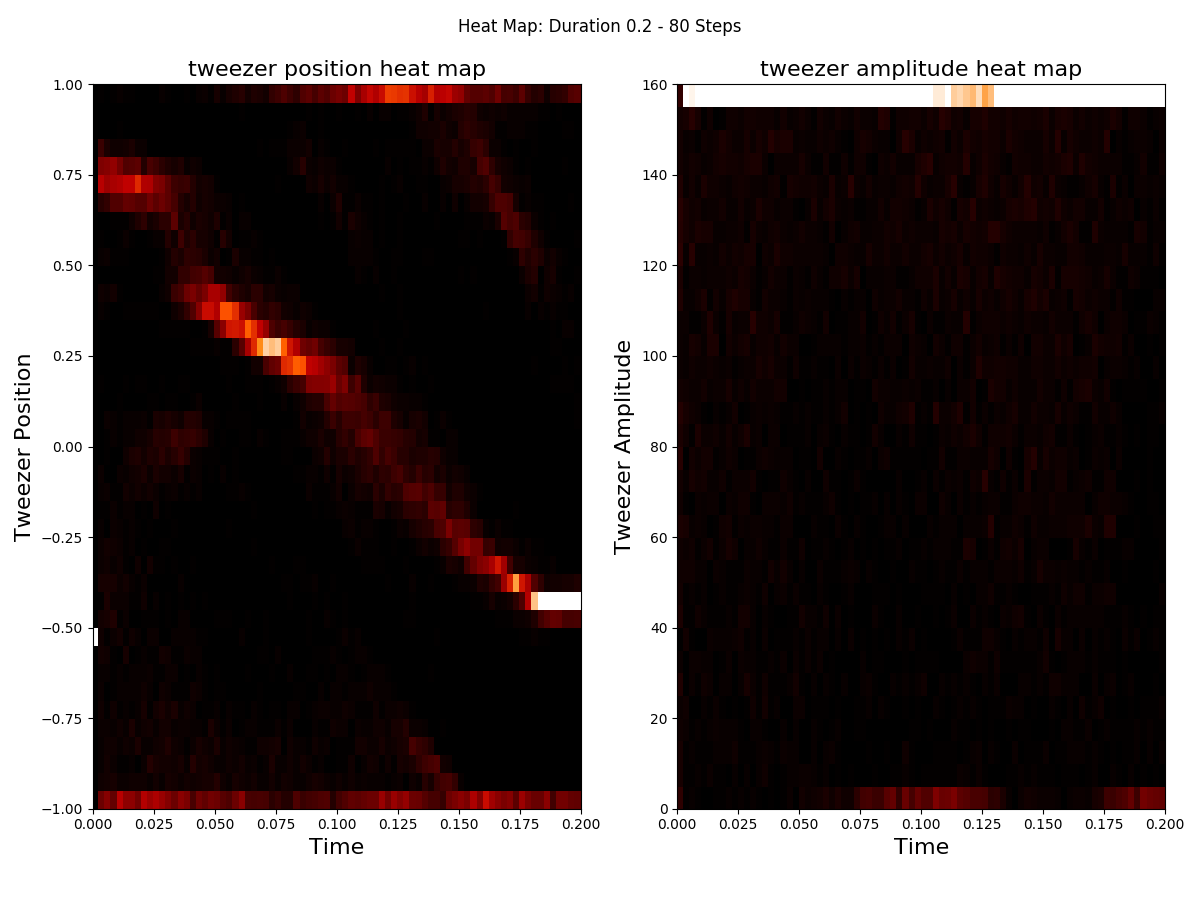}
  \caption{Grape Position and Amplitude Heat Map for duration 0.20, using 80 steps}
  \label{fig:hf}\end{subfigure}
\end{tabular}
\caption{Heat Maps for GRAPE positions and amplitudes for different durations. The heat maps are generated from the 200 GRAPE runs with best fidelity among 2000 runs.  
}
\label{tab:heat_maps}
\end{table}

\subsection{Krotov Algorithm}
To implement the Krotov algorithm, we use the already existing implementation of GRAPE for computing Hamilton's and unitaries and base our Tensorflow implementation on that. We do not use automated differentiation for the Krotov algorithm as it turns out to be simple to write out the necessary gradient computations.  The number of steps, $N$, for a given duration, $T$, is set, as for stochastic ascent and GRAPE, such that $T/N = 0.0025$. For each duration $T$ tested we ran several Krotov invocations from completely uniform random starting positions and amplitudes.
As for stochastic ascent and GRAPE, in order to save time, we ran the main bulk of experiments using imprecise computation by simply setting a low value for the grid size $h_g$ and running each experiment only a limited amount of iterations. We ran 1000 experiments starting with,  $h_g=32$ and ran it for at most 1500 iterations or until it had stopped improving.  Then the best protocol found like this is considered again, we increase the grid size back by doubling and restarting the iteration, until we hit a grid size of 512. 
As for GRAPE, it is important to handle the learning rates properly, and as for GRAPE, we again use the ADAM \cite{kingma2014adam} algorithm for that , using a separate optimizer for positions and amplitudes.
For the positions, we start with an initial learning rate of 0.01 and reduce it during the algorithm when the fidelity stops improving. Similarly, for the amplitudes, except that we start with a learning rate factor of 0.1. The reason for making this learning rate higher for amplitudes is simply that the gradient of the positions include the amplitudes as a scale factor, making them much larger than the position derivatives in absolute value. The same thing could have been added to our GRAPE implementation.
It takes a few hours on a standard work station to run all 1000 random starts for a given duration and optimizing the best found protocol from there takes even less. Again, for a full test one should really test different optimization methods and very their hyper-parameters. To ensure the tweezer starts as in the game we always fix the first position of the protocol to $-0.55$ and never update it in the algorithm.

\subsection{Actual Code}
Code for running the algorithms and notebooks that shows how to run the algorithms by example can be found at: \url{https://github.com/gronlund/QuantumMoves}.

\end{document}